\documentclass[sigconf,natbib=false,10pt]{acmart}
\renewcommand\footnotetextcopyrightpermission[1]{} 
\acmConference[]{}
\copyrightyear{}

\AtBeginDocument{%
  }

\newif\ifcomment
\commentfalse

\ifcomment
    \newcounter{ALNumberOfComments}
    \stepcounter{ALNumberOfComments}
    \newcommand{\ael}[1]{\textcolor{magenta}{\small \bf [AL\#\arabic{ALNumberOfComments}\stepcounter{ALNumberOfComments}: #1]}}
    
    \newcounter{SBNumberOfComments}
    \stepcounter{SBNumberOfComments}
    \newcommand{\ssba}[1]{\textcolor{red}{\small \bf [SB\#\arabic{SBNumberOfComments}\stepcounter{SBNumberOfComments}: #1]}}
    
    \newcounter{PANumberOfComments}
    \stepcounter{PANumberOfComments}
    \newcommand{\paul}[1]{\textcolor{blue}{\small \bf [PA\#\arabic{PANumberOfComments}\stepcounter{PANumberOfComments}: #1]}}
    
    \newcounter{JSVNumberOfComments}
    \stepcounter{JSVNumberOfComments}
    \newcommand{\jsv}[1]{\textcolor{purple}{\small \bf [JSV\#\arabic{JSVNumberOfComments}\stepcounter{JSVNumberOfComments}: #1]}}
\else
    \newcommand\ael[1]{}
    \newcommand\paul[1]{}
    \newcommand\ssba[1]{}
    \newcommand\jsv[1]{}
\fi

\usepackage{xspace}
\usepackage[table]{xcolor}
\usepackage{tabularx,colortbl,multirow}
\usepackage{booktabs}     
\usepackage{threeparttable}
 \usepackage{makecell}
 
 \usepackage{graphicx}  
\usepackage{amsmath,  bm} 
 \usepackage{float}
 \usepackage{tikz}
\usepackage{graphicx}
\usepackage{booktabs}
\usepackage{multirow}
\usepackage{adjustbox}
\usepackage[skip=0.2pt]{caption}
\usepackage{subcaption}
\usepackage[printonlyused]{acronym}
\acrodef{RAN}{Radio Access Network}
\acrodef{MNO}{Mobile Network Operator}
\acrodef{CDF}{Cumulative Distribution Function}
\acrodef{RSRP}{Reference Signal Received Power}
\acrodef{SINR}{Signal to Interference and Noise Ratio}
\acrodef{MIMO}{Multiple-Input and Multiple Output}
\acrodef{RIS}{Reconfigurable Intelligent Surfaces}
\acrodef{AI}{Artificial Intelligence}
\acrodef{PL}{Path Loss}
\acrodef{RSS}{Received Signal Strength}
\acrodef{DL}{Deep Learning}

\usepackage[most]{tcolorbox}
\tcbset{
  maintakeaway/.style={
    colback=gray!10,
    colframe=gray!50,
    boxrule=0.5pt,
    arc=1mm,
    left=1mm, right=1mm, top=1mm, bottom=1mm,
    before skip=4pt, after skip=4pt,
    width=\columnwidth,
    enhanced
  }
}

\usepackage{overpic} 

\newcommand{\eg}{\textit{e.g.},\xspace}
\newcommand{\ie}{\textit{i.e.},\xspace}

\RequirePackage[
  datamodel=acmdatamodel,
  style=acmnumeric,
  ]{biblatex}

\begin{document}

\title[To Differentiate or To Deep Learn?]{Radio Propagation Modelling: To Differentiate \\ or To Deep Learn, That Is The Question}
\thispagestyle{plain}

\author{Stefanos Bakirtzis}
\email{ssb45@cam.ac.uk}
\affiliation{%
  \institution{University of Cambirdge}
  \city{Cambridge}
  \country{United Kingdom}
}

\author{Paul Almasan}
\affiliation{%
  \institution{Telefonica Research}
  \city{Barcelona}
  \country{Spain}}

\author{José~Suárez-Varela}
\affiliation{%
  \institution{Telefonica Research}
  \city{Madrid}
  \country{Spain}
}

\author{Gabriel O. Ferreira}
\affiliation{%
  \institution{IEIIT CNR Institute}
  \city{Turin}
  \country{Italy}}

\author{Michail Kalntis}
\affiliation{%
  \institution{Delft University of Technology}
  \city{Delft}
  \country{Netherlands}}

\author{André Felipe Zanella}
\affiliation{%
  \institution{Telefonica Research}
  \city{Barcelona}
  \country{Spain}}

\author{Ian Wassell}
\affiliation{%
  \institution{University of Cambridge}
  \city{Cambridge}
  \country{United Kingdom}}

\author{Andra Lutu}
\affiliation{%
 \institution{Telefonica Research}
 \city{Madrid} 
 \country{Spain}}
 
\renewcommand{\shortauthors}{Bakirtzis et al.}

\begin{abstract}

Differentiable ray tracing has recently challenged the status quo in radio propagation modelling and digital twinning. 
Promising unprecedented speed and the ability to learn from real-world data, it offers a real alternative to conventional deep learning (DL) models. 
However, no experimental evaluation on production-grade networks has yet validated its assumed scalability or practical benefits. 
This leaves mobile network operators (MNOs) and the research community without clear guidance on its applicability. 
In this paper, we fill this gap by employing both \emph{differentiable ray tracing} and \emph{DL models} to emulate radio coverage using extensive real-world data collected from the network of a major MNO, covering 13 cities and more than 10,000 antennas. 
Our results show that, while differentiable ray-tracing simulators have contributed to reducing the efficiency–accuracy gap, they struggle to generalize from real-world data at a large scale, and they remain unsuitable for real-time applications. 
In contrast, DL models demonstrate higher accuracy and faster adaptation than differentiable ray-tracing simulators across urban, suburban, and rural deployments, achieving accuracy gains of up to 3 dB.\paul{What does it mean "gains up to 2-4 dB"? Do you mean that the error is a low as 2-4 dB?}
Our experimental results aim to provide timely insights into a fundamental open question with direct implications on the wireless ecosystem and future research.  
\end{abstract} 
\maketitle


\vspace{-2mm}
\section{Introduction}

Radio propagation models~\cite{Hata, RadioPro_Urban_1978, Selecive_Fading_1979, RadioPro_City_Uni_1988, Statistical_Model_}, developed to predict wireless signal behavior in real environments, have been central to network planning and deployment, and remain fundamental to enabling advanced technologies like \ac{MIMO}, beamforming, \ac{RIS}, localization, and power control \cite{RT_Principles_applications, Radio_Propagation_Models_5G, Empowering_DL_Prop_Models}.
As wireless networks grew more complex, traditional models like Hata~\cite{Hata} became inadequate. Physics-based approaches such as ray-tracing improve accuracy, while remaining computationally demanding.
  
In the era of \ac{AI}, the research community has been proposing solutions based on \ac{DL} to overcome this core limitation~\cite{AI_Prop_Survey, AI_Prop_Editiorial, Aris_AI_Prop, Cagkan_Overview}; a direction also supported by major players in the industry~\cite{Ranplan_AI_Prop, Infovista_AI_Prop, iBwave_AI_Prop}.
The vast preponderance of \ac{DL} approaches relies on applying neural transformations to some input data and converting it into a target quantity, such as the \ac{PL} or \ac{RSS} radio maps. These include models based on simple multilayer perceptron (MLP) \cite{Azpilicueta_Neural_RL, Macro_ANN, Popescu2_ANN_2025} and convolutional neural networks (CNNs)~\cite{SIGKDD_Cellular, Radio_U_Net, Molisch_CNN, FadeNET, Radio_Prop_CNN, EM_DeepRay, Marco_ICASSP, DL_Indoor, Kehai_Outdoors, Indoor_PL} or, more recently, transformers \cite{Rafayel_Trasnformers, Transofrmer_Urban, viet2024spatial}, graph neural networks~\cite{Graphs_TAP, Graphs_VET, Radio_Gat, Graph_Reconsutr}, and diffusion models~\cite{Diffusion_Prop,RD_diffusion, Kehai_diffusion, Radio_Diff}, as well as those incorporating advances in computer vision~\cite{Nerf_Paper, Gaussian_Splatting_Paper}.  The input data or the transformation type may vary, \eg applying convolutional, attention-based, message-passing on indoor/outdoor physics-based features; however, the essence of these approaches remains the same: DL-based models are used merely as black-boxes that learn from data without considering the governing physical principles of the underlying system. On the other hand, approaches trying to integrate the laws of physics into AI-driven models ~\cite{PINN_EM_Integral_Eq,Diffusion_Physics_Informed,PINNS_RF_Body} or exploit computer vision techniques \cite{Nerf2, Newrf, WRF_GS_Splatting, RF-3DGS_Gaussian_Splatting, WiNERT} require collecting extensive data for each scene. More importantly,  their predictions are not transferable to unknown environments as they learn to retrieve solutions under a specific physical problem parametrization.  

This surge in the development of DL-driven solutions, along with their successes and shortcomings, comes with a concealed and unanswered, yet 
fundamental research question (RQ):


\begin{tcolorbox}[maintakeaway]
\textit{\textbf{RQ}}: Is the extensive use of deep learning 
justified, or has the research community been taken over by the AI hype and its blind adoption?
\end{tcolorbox}


Importantly, this question is not only relevant to the field of radio propagation and wireless communications, but also to scientific research \textit{in toto}. Consequently, we argue that it is paramount to critically and empirically explore the role and degree of integration of AI in future technological systems and their components. 

In this paper, we study whether it is necessary to strive to develop intricate DL-based propagation models, or if it is enough to revise and improve the implementation of legacy and well-established methods with the potential partial integration of AI.  
 For radio propagation modelling, our question becomes particularly critical in light of recent advances reporting a ``lightning-fast" and differentiable ray-tracing implementation~\cite{Nvidia_Lighting_Fast_RT}, namely  Sionna RT~\cite{Sionna_Globecom, Sionna_General} (hereafter, we refer to this ray tracing package simply as Sionna).
This conceptualization can entirely change the research landscape and shift the focus of the mobile and networking community, since ($i$) Sionna is claimed to match or even exceed the speed of AI-driven approaches~\cite{Nvidia_Lighting_Fast_RT}, and ($ii$) due to its differentiable nature, it can assimilate knowledge from measured data and demonstrate real-world accuracy levels~\cite{SIonna_Learning}. 
Apparently, this explicitly diminishes the longstanding argument against legacy radio propagation solvers, and to extent, the motivation behind the relentless development of DL-based radio propagation models. However, to the best of our knowledge, \textit{we lack a large-scale experimental validation or evidence to substantiate these claims to date}, and support either research direction. 
Hence, while one side of the research community propels innovation through (brute-force) exploitation of AI, another advocates the return to foundational computational methods, selectively augmented by the capabilities of AI and differentiability. The question of which perspective is ultimately correct still remains open.

Our paper provides a timely perspective to this debate, documenting our experience with state-of-the-art DL-based solutions and differentiable ray-tracing in a nationwide commercial radio network deployment. To this end, we leverage a vast dataset that blends: ($i$) network topology inventory of 10,000+ antennas from the \ac{RAN} of a commercial countrywide \ac{MNO}, ($ii$) 6-months of real-world crowdsourced radio coverage measurements from the \ac{MNO}'s end-users connected to these antennas, and ($iii$) geographic information for the areas served by the antennas.
Our work yields the following contributions:

\setlength{\leftmargini}{0pt}
\begin{itemize}
\item \textbf{We present the first large-scale experimental study on the scalability of differentiable ray-tracing in commercial radio networks.
}
Simulations on the commercial RAN deployment show that differentiable ray-tracing improves fidelity and adaptability in complex environments, but its computational cost limits practical adoption, raising concerns for \acp{MNO} and underscoring the need to address these challenges.

\item \textbf{We conduct a thorough comparative analysis between differentiable ray-tracing and state-of-the-art DL models, assessing their potential to generate radio maps that align with real-world data
.} 
We demonstrate that, 
amid recent advances in radio propagation modelling, 
the use of differentiable ray-tracing surrogates appears limited. 
In contrast, DL models trained on massive real-world data generalize effectively and operate as full-fledged radio map solvers. Most importantly, \ac{DL} consistently outperforms differentiable ray-tracing in scalability, accuracy, and efficiency, making it the leading approach for real-time, high-fidelity radio map generation 
at large scale.


\item \textbf{We quantify the impact of inaccuracies in both differentiable ray-tracing and \ac{DL} models on downstream tasks that rely on the predicted radio maps as input.
}
Specifically, we tackle ($i$) an advanced optimization model for green networking~\cite{Gabrielle_Paper} and ($ii$) a cutting-edge technique for 
mobility management optimization~\cite{kalntis_infocom25}. 
Our results indicate that, unlike differentiable ray tracing, radio maps generated by \ac{DL} models yield results comparable to those obtained using real-world data across both tasks. 
This confirms the potential of such models as reliable tools for \acp{MNO} to build digital replicas of their RANs, while also highlighting the need for future research to narrow the accuracy gap between DL-generated and real-world coverage maps.

\end{itemize}


\section{Related Work}
\label{sec:related}

 
In this section, we offer an overview of the main classes of radio propagation models to date and discuss their strengths and weaknesses.

\begin{figure*}[!t]
\centering
    {\includegraphics[scale = 0.525]{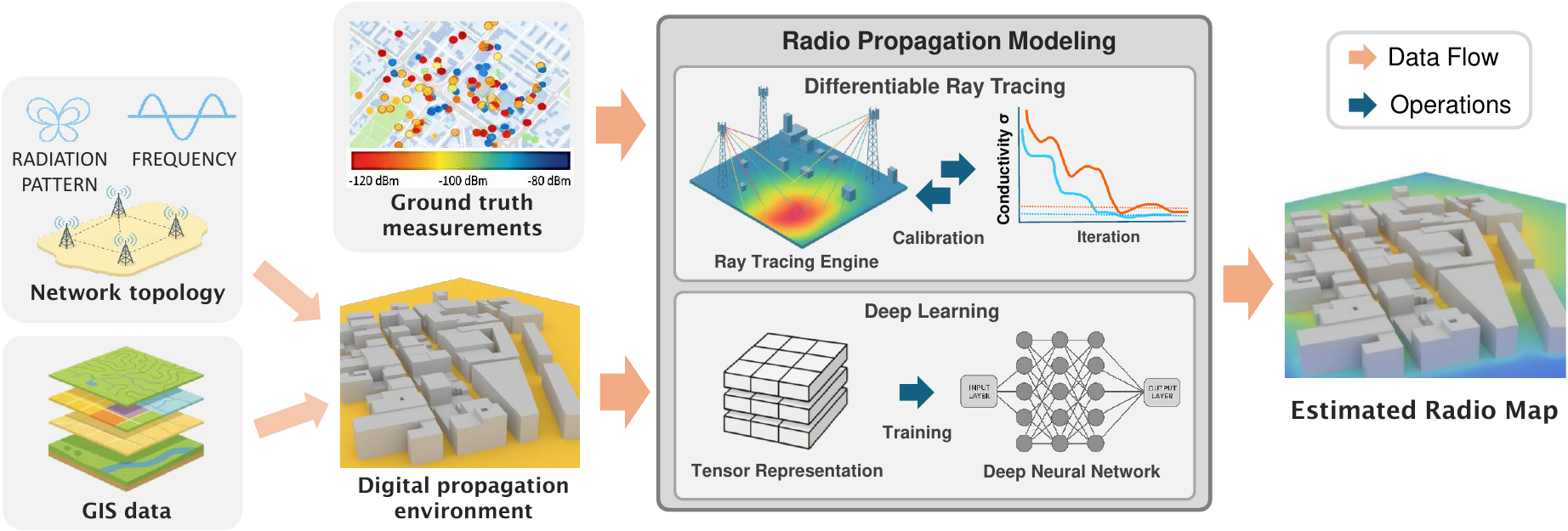}} 
    \vspace{0.2cm}
\caption{Schematic representation of the developed radio propagation modelling framework.}
\vspace{-0.2cm}
\label{fig:General_Pipeline}
\end{figure*}

\noindent \textbf{Empirical models}.  From the deployment of the first communication systems, researchers, along with network and radio engineers, aspired to establish methods and models that could quantify how wireless signals propagate in the real-world environments where transmitters operate~\cite{Hata, RadioPro_Urban_1978, Selecive_Fading_1979, RadioPro_City_Uni_1988, Statistical_Model_}. Empirical models rely on statistics and extensive measurements to predict radio wave propagation, and they are characterized by their simplicity and low computational cost. The Hata model~\cite{Hata} is one of the most well-known methods derived from field measurements and it has an applicable frequency range from 150-1500 MHz. 
More recently, interpolation-based methods were proposed to leverage measurements to reconstruct radio maps~\cite{9893802, 10.1145/3308558.3313726, 6150651}. 
Despite their benefits, empirical models are no longer viable solutions and cannot meet the performance requirements of modern networks: they exhibit large prediction errors in complex propagation environments, are constrained to site-specific parametrization or calibration environments, and their performance varies significantly across different geographical settings.  



\noindent\textbf{Deterministic  models}. To engineer more realistic, site-specific, and physics-consistent methods, approximate solutions to Maxwell equations were proposed. Among them, ray-tracing or ray launching \cite{Ray_Tracing_Networks, Ray_Tracing_Techniques, Ray_Launching} has been widely adopted by radio practitioners and many commercial network planning simulation tools~\cite{atoll, wirelessinside, Ranplan} due to its potential to emulate radio wave trajectories and multipath effects accurately and comparatively faster than other full-wave methods. 
However, ray-tracing can still be time-consuming, taking seconds to several minutes per simulation \cite{Aris_Planning, WithRay}. 

Recently, Nvidia propelled a paradigm shift in ray-tracing, introducing Sionna~\cite{Nvidia_Lighting_Fast_RT, Sionna_Globecom}, an open-source differentiable ray-tracing model. 
Sionna achieves low simulation speeds (see Section~\ref{sec:models-comparison}), and its ray-tracing outputs can be differentiated with respect to system parameters, \eg antenna configuration or material electromagnetic properties.




\noindent \textbf{Deep Learning models.} Recent AI-based methods for radio map prediction span from traditional neural networks~\cite{Azpilicueta_Neural_RL,Macro_ANN,Popescu2_ANN_2025,SIGKDD_Cellular,Radio_U_Net,Molisch_CNN,FadeNET,Radio_Prop_CNN,EM_DeepRay,Marco_ICASSP,DL_Indoor,Kehai_Outdoors,Indoor_PL}, to emerging architectures including transformers~\cite{Rafayel_Trasnformers,Transofrmer_Urban,viet2024spatial}, graph neural networks (GNNs)~\cite{Graphs_TAP,Graphs_VET,Radio_Gat,Graph_Reconsutr}, and diffusion models~\cite{Diffusion_Prop,RD_diffusion,Kehai_diffusion,Radio_Diff}. These approaches typically learn a mapping from features (\eg city layout and materials) to radio maps, functioning largely as black-box regressors without explicit modelling of physical propagation mechanisms. Notably, most of these works are deployed with synthetic data and very few probe the potential of these models learning from real-world measurements, highlighting a significant gap in practical evaluation and deployment. Moreover, some promising approaches based on advanced models, such as diffusion models, leave open the challenge of adapting these methods to learn from sparse measurements~\cite{Radio_Diff}.

More recent approaches strive to exploit the latest advances in computer vision, used to describe optical fields, \eg neural radiance field (NeRF) \cite{Nerf_Paper} or three-dimensional (3D) Gaussian splitting (GS) \cite{Gaussian_Splatting_Paper}, and tailor them to portray the interactions between the physical environment and radio signals \cite{Nerf2, Newrf, WRF_GS_Splatting, RF-3DGS_Gaussian_Splatting, WiNERT}. These models are typically trained to infer detailed channel information, \eg full spatial radio spectrum, but they entail rendering the entire propagation environment. That can pose significant scalability issues in realistic large-scale urban deployments, and therefore, as of now, such models have been tested solely in simplified indoor environments \cite{Nerf2, Newrf, WRF_GS_Splatting, RF-3DGS_Gaussian_Splatting, WiNERT}.
Likewise, physics-informed neural networks~\cite{PINN_EM_Integral_Eq,Diffusion_Physics_Informed,PINNS_RF_Body} attempt to embed electromagnetic principles into the learning process; but they are limited by problem specific parametrization, restricting their transferability to new environments. Consequently, due to these generalization constraints, such approaches are currently impractical for deployment 
in large-scale real-world networks.

 


\section{Radio Propagation Modelling Framework}
\label{sec:radio-prop-framework}

In this section, we describe the framework we have developed to evaluate state-of-the-art differentiable ray-tracing and DL models at a large scale on a wide collection of datasets from a commercial MNO. Figure~\ref{fig:General_Pipeline} shows an overview of the developed framework. The following sections describe each component of this framework in detail, including the datasets used and the different modules we implemented.

\subsection{Datasets}\label{sec:datasets}

We use three main data sources for our real-world measurement study. Integrating these multi-modal data from various sources allows us to construct a realistic digital twin of the radio propagation environment, as well as run large-scale radio propagation simulations through differentiable ray-tracing and DL models. 

(1) \textbf{Network topology}: a database with the exact locations and configuration parameters of 10,000+ antennas deployed in urban, suburban and rural areas. This includes the antenna's operating frequency, $f$, the  3D antenna radiation pattern, $G(\phi, \theta)$, the steering angles in the horizontal and vertical directions (\ie their azimuth and tilt, respectively), the transmitting power $P_{Tx}$, and additional hardware losses $L_H$ (\eg the feeder cable and connector losses). 
 
(2) \textbf{Ground truth measurements}: For each antenna, we collect crowdsourced data from a diverse set of end-user devices connected to it, conveying large-scale insights about the signal strength distribution around the antenna. For this, we leverage a proprietary app that the MNO 
disseminates to its end-user population, with the end-goal of improving the quality of service and end-user experience.   
Specifically, each measurement record comprises: ($i$) the \ac{RSRP} and the \ac{SINR}, ($ii$) the latitude and longitude of the user when the recording occurred, ($iii$) a flag indicating whether a user is indoors or outdoors, and ($iv$) an estimate of the accuracy of the user location. The crowdsourced data were collected over a six-month period, starting from the 1\textsuperscript{st} of April 2024 until the end of September 2024. 

(3) \textbf{GIS data}: We use proprietary quality-checked geographic information system (GIS) databases\footnote{We are using a commercial provider for GIS data.}  comprising: ($i$) building-level data including features such as the building height and construction material,  ($ii$) reliable land use information with 23 distinct land-use types (\eg woodlands,  business parks, industrial or residential zones, retail parks), and ($iii$) a detailed representation of the road network. 

 For our analysis, we select the commercial antennas for which we collected more than 1,000 measurement records during the 6-month campaign period.
This results in considering approximately 10,000 antennas located in 13 distinct urban, suburban, and rural areas of the studied country\footnote{Throughout the paper, we refer to these areas as Area 1 to 13}, which have a total of $\approx$300,000,000 associated measurements. These areas collectively occupy a region of almost 3,000 square meters, with an average area size equal to approximately 200 square meters.  We note that the number of measured data entries is evaluated after removing measurement samples found in indoor environments or having an estimate of the position accuracy higher than 10 meters.

\begin{figure}[t]
\centering

\begin{subfigure}[b]{0.48\columnwidth}
  \centering
  \begin{tikzpicture}
    \node[anchor=south west, inner sep=0] (main) at (0,0)
      {\includegraphics[width=\columnwidth,trim=0 0 0 0,clip]{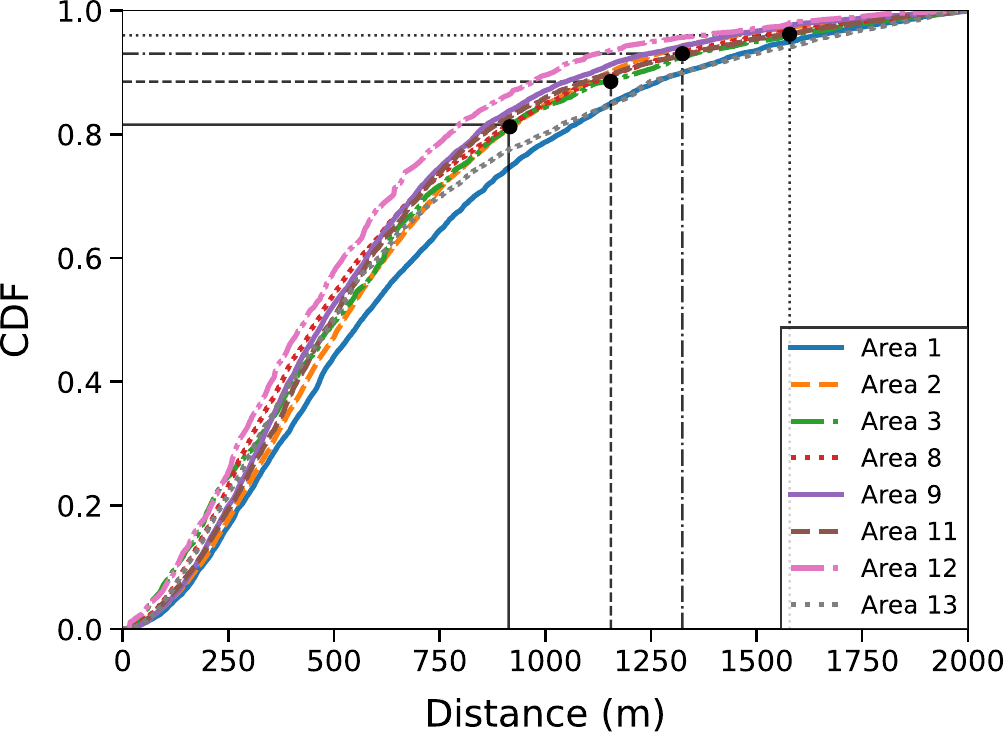}};
    \node[anchor=north east, xshift=125pt, yshift=8pt] at (main.north east)
      {\includegraphics[width=1.8\columnwidth,trim=0 0 0 0,clip]{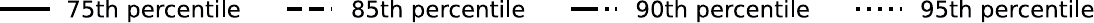}};
  \end{tikzpicture}
    \vspace{-3mm}
  \caption{Group 1}
  \label{fig:group1}
\end{subfigure}
\hfill
\begin{subfigure}[b]{0.48\columnwidth}
  \centering
  \includegraphics[width=\columnwidth,trim=0 0 0 0,clip]{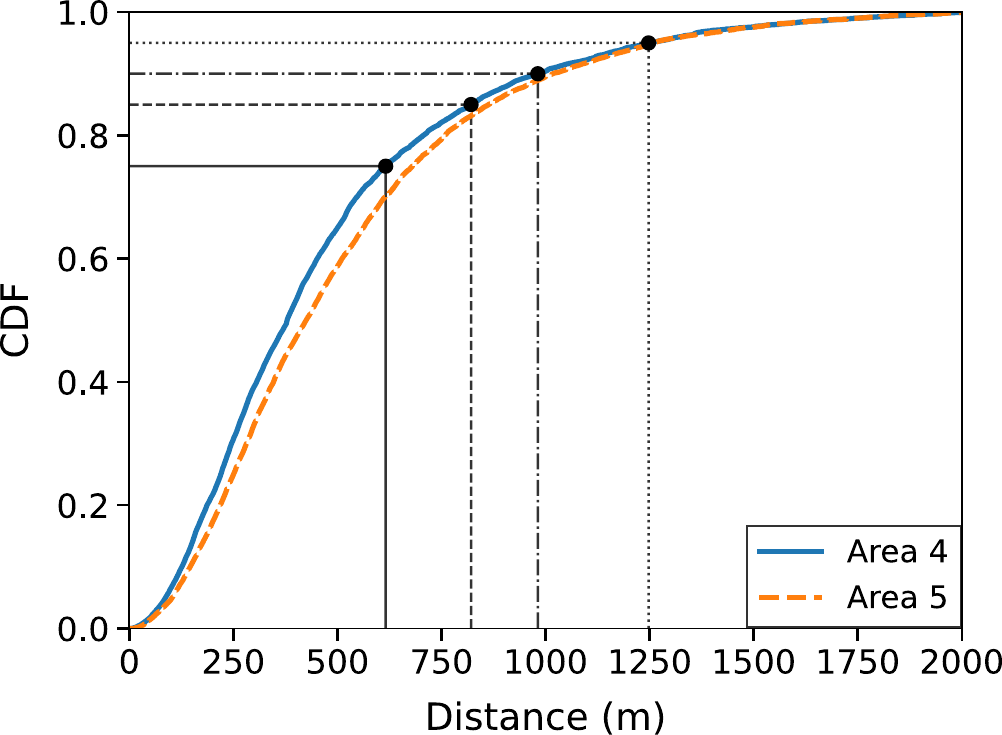} 
  \caption{Group 2}
  \label{fig:group2}
\end{subfigure}

\vspace{-2pt}

\begin{subfigure}[b]{0.48\columnwidth}
  \centering
  \includegraphics[width=\columnwidth,trim=0 0 0 0,clip]{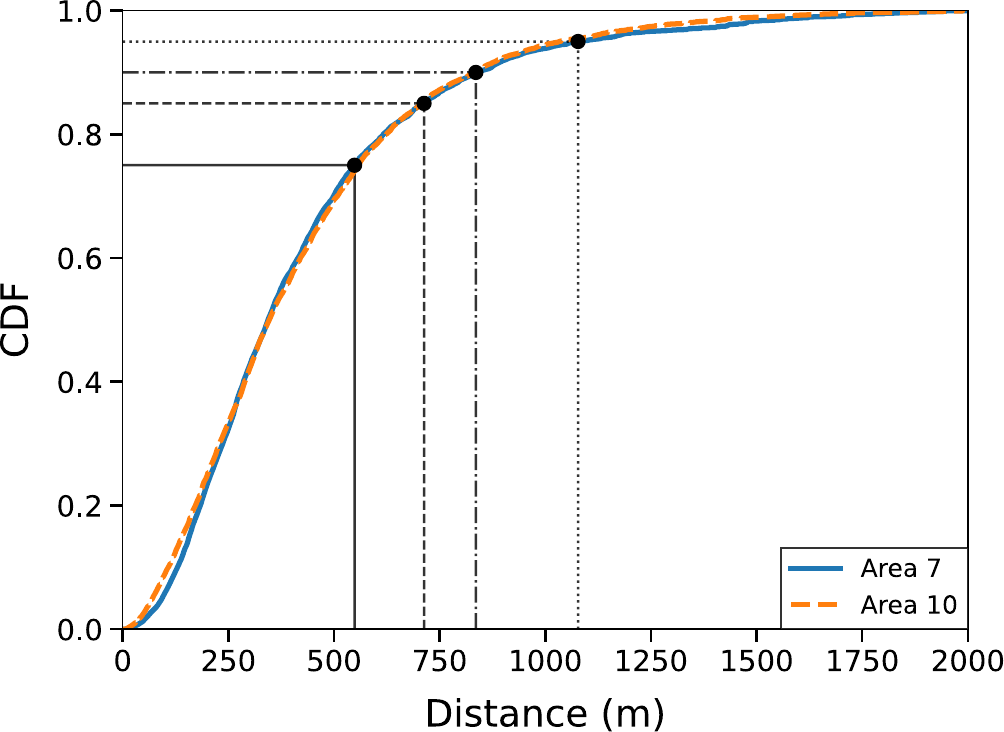} 
  \caption{Group 3}
  \label{fig:group3}
\end{subfigure}
\hfill
\begin{subfigure}[b]{0.48\columnwidth}
  \centering
  \includegraphics[width=\columnwidth,trim=0 0 0 0,clip]{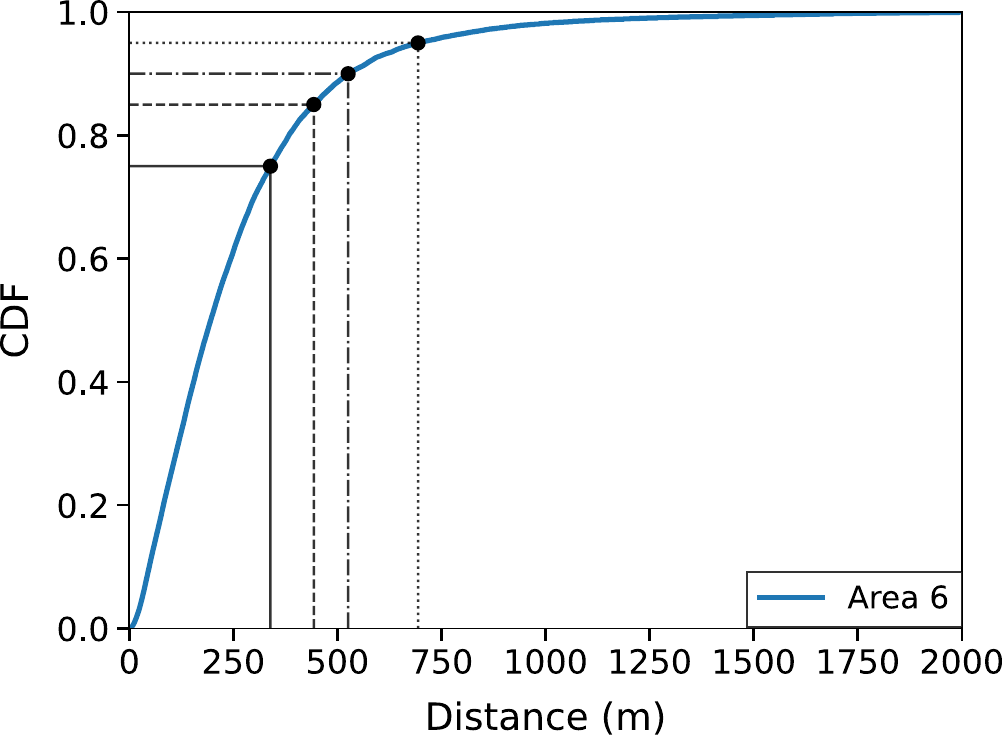} 
  \caption{Group 4}
  \label{fig:group4}
\end{subfigure}

\caption{CDFs of measurement-to-antenna distance across the different areas of study.}
\label{fig:Distance_CDFs}
\end{figure}


\subsection{Ground-Truth Radio Maps} 
\label{sec:Radio_Maps}

To assess whether differentiable ray-tracing and \ac{DL}-based propagation models can faithfully estimate real-world RAN coverage, we build the RSRP coverage map for each antenna using the corresponding (ground truth) coverage measurements. 
To this end, we rasterize the area around the antenna by dividing it into a grid of equal-sized cells (with the antenna always located at the centre of the grid). Then, we associate with each grid cell the measurement samples (based on latitude-longitude filtering) and generate our ground truth maps following the process we describe next. 

Figure~\ref{fig:Distance_CDFs} shows the \ac{CDF} of the measurement-to-antenna distance, reflecting the percentage of measurement samples that fall below a given distance from the serving antenna. Specifically, we classify Areas 1 to 13 into four groups\footnote{The group labels are used solely to distinguish areas with similar distance–RSRP percentage distributions and do not imply any other type of similarity and correlation between the areas considered.} according to the similarity of their aggregated RSRP percentage distribution over distance. 

We find that the distance distribution around the serving antennas varies for the different groups, whilst areas within the same group exhibit a similar profile in terms of the measured coverage distance around the antennas. \jsv{@Stefanos, pls check this paragraph provides accurate definitions}

To ensure that the generated RSRP coverage maps remain representative, for each area we determine the geographic extent of its antennas, defined as the maximum distance from the serving antenna such that $90\% $ of the available measurement data fall within the rasterized grid. As shown in Figure~\ref{fig:Distance_CDFs}, in areas of Group 1 (8 out of the 13 Areas), this threshold is reached at approximately 1,250 meters from the antenna. For the two other groups, referred to as Groups 2 and 3, $90\%$ of the measurement data lie within a radius of approximately 1,000 meters and 800 meters, respectively. Lastly, the area encompassing the capital city of the studied country (Area 6), which has a higher deployment density, exhibits a threshold slightly above 500 meters.

As discussed in Section \ref{sec:Radio_Maps}, our DL models operate with fixed-size tensors of $512 \times 512$. To facilitate the DL training pipeline, the maximum distances we consider are 1280, 1024, 768, and 512 meters away from the antenna for Group 1, 2, 3, and 4, respectively. Accordingly, we set spatial resolutions to 5, 4, 3, and 2 meters per grid cell to align with the geographic extent of antennas within each group. Particularly, the total grid size spans twice the maximum distance, as this value represents the radius from the antenna, hence defining the full coverage area. Interestingly, most previous work considers substantially smaller geographic areas, \eg $128 \times 128$ or  $256 \times 256$ meters, which appears to be unsuitable for representing realistic coverage regions. 

The emerging raster grid is explicitly mapped to a geographic coordinate system, with grid cell centres corresponding to distinct latitude–longitude pairs, generated as evenly spaced sequences around the antenna location. That allows mapping each measurement from our dataset onto a discrete grid cell by computing its grid indices based on its offset from the grid origin and cell size. Hence, for each grid cell, we can aggregate all measurements falling within its boundaries to compute the median RSRP values, and generate our ground truth RSRP coverage maps. For convenience, any grid cells without any measurements falling within their boundary are filled with zeros. 


\subsection{Digital Propagation Environment} \label{sec:Dig_Prop_Env}

 To create a digital representation of the radio propagation environment, similar to those shown in Figure~\ref{fig:General_Pipeline}, we exploit our proprietary GIS database to systematically extract 3D representations from geospatial buildings data. Specifically, we represent each building in our database as a polygon consisting of (latitude, longitude) pairs that defines a 2D footprint on the ground. We convert the 2D building footprint into a triangular mesh using a Delaunay tessellation, ensuring a well-defined mesh topology. 
 This mesh is then extruded vertically according to the building’s height, resulting in a closed 3D solid model with capped tops. 
 Each mesh is saved as a PLY file and incorporated into the XML scene, with the material type set according to our database: concrete, brick, or glass. This approach allows the XML scene to faithfully represent the geometric and material characteristics of the propagation environment, enabling accurate simulations of signal propagation. 

\subsubsection{\textbf{Ray tracing network representation}}\label{sec:Digital_Net_Rep} 

The scene format described above is suitable for simulation in our ray-tracing engine (Sionna). With the digital radio propagation environment designed, we can set up a digital equivalent of the operating RAN, exploiting the information provided by the MNO. As mentioned in Section~\ref{sec:Radio_Maps}, the grid is centred around the antenna's latitude and longitude; hence, the antenna is always placed at the centre of the scene, with its height, $z_{Tx}$, set according to our database records. Similarly, we are able to define its operating frequency, bandwidth, transmitting power, and 3D antenna radiation pattern, as well as its steering angle in the horizontal and vertical dimensions, respectively. This process creates a digital twin of the production-grade network that can be used to conduct ray-tracing simulations to emulate signal propagation around an antenna and benchmark it with the generated RSRP radio maps. 

\subsubsection{\textbf{DL-based tensor representation}}\label{sec:DL-tensor-representation}
 A more simplified representation can be used to portray the GIS information in a format suitable for DL-based propagation models. Indeed, such models operate over multi-dimensional tensors; thus, one solely needs to transform vector geometries into a rasterized grid format and create a layered 2D tensor representation. To this end, we first compute an affine transformation that defines the precise spatial relationship between the rasterized grid cell indices and real-world coordinates. Then, each GIS database entry is paired with its associated feature value, \eg building height or categorical land use value, forming a list of (geometry, value) tuples. This list is used along with the affine transformation to populate a 2D grid, assigning the feature values to cells whose centres fall within the input geometries. The result is a three-layer 2D tensor representation comprising the following grids: ($i$)~Building height, indicating both the presence of a building in a grid cell and its corresponding height (set to 0 if no building is present); ($ii$) Categorical land use, with classes ranging from 0 to 23, representing the land-use categories defined in our GIS database; and ($iii$) Binary road information, denoting whether a grid cell overlaps with a road.
 
To incorporate antenna-related data into DL models, we apply the following steps. First, we consider a $(x, y)$ Cartesian coordinate system with its origin located at the antenna position, and with subsequent points spaced apart with the resolution assigned to each area, as per Section~\ref{sec:Radio_Maps}. That allows us to calculate the 3D distance between the transmitting antenna (Tx) and each point of the grid, assuming that the receiver (Rx) is located at $z_{R_x} = 1.5$ meters above the ground; $d = \sqrt{ \allowbreak (x_{T_{x}} - x_{R_{x}} )^2 
  + \allowbreak (y_{T_{x}} - y_{R_{x}} )^2 
  + \allowbreak (z_{T_{x}} - z_{R_{x}} )^2}$. Then, we compute the azimuth, $\phi$, and elevation angles, $\theta$, respectively, between the Tx and Rx for all the points of the receiving plane. To do that, a 2D rotation matrix is applied to the horizontal coordinates to align the coordinate system with the antenna's azimuth orientation, ensuring proper directional alignment regardless of the antenna's position within the grid.   Subsequently, a 3D rotation is performed specifically for the elevation (tilt) component around the y-axis, maintaining the azimuth alignment while incorporating the antenna downtilt. This dual-transformation approach results to a new $(x', y', z')$ coordinate system, yielding accurate coordinates for each grid point relative to the antenna's boresight direction. The azimuth and elevation angels are computed as:  $ \phi = Arctan( \frac{y'_{R_{x}} - y'_{T_{x} } } {x'_{R_{x}} - x'_{T_{x}}}  )$, and $\theta = Arcsin(  \frac{z'_{R_{x}} -  z'_{T_{x}}   } { d } )$, respectively. This results in three 2D tensors, $\mathbf{D}$,  $\mathbf{\Phi}$, $\mathbf{\Theta}$, including the distance, azimuth, and elevation angle from the antenna, respectively.

  These tensors are used to derive a single compact tensor representation, $\mathbf{L}$,  that portrays the impact of the free space path loss (FSPL), antenna radiation pattern, and transmitting power on the received signal strength. Specifically, for a given 3D antenna radiation pattern,  $G(\phi, 
  \theta)$, then $\mathbf{L} =  20 log_{10} \left ( \frac{4 \pi \mathbf{D} f}{ c_o} \right ) \allowbreak + G(\mathbf{\Phi} ,\mathbf{\Theta}) + P_{Tx} + L_{H}$. Note that in the free space, $L$ is almost identical to Sionna's outputs. As such, this channel serves as a baseline upon which a DL model can be trained to reconstruct the target result.



\section{To Deep Learn or Not?}
\label{sec:models-comparison}

To answer the fundamental question motivating this work, we test the performance over a collection of five state-of-the-art DL models. 
In addition, we consider a spectrum of comprehensive experiments for an extensive and fair comparison between DL-based models and differentiable ray-tracing:

 \textbf{\texttt{Ray-tracing to Ray-tracing (R2R)}}: We train the DL models with synthetic radio maps generated using Sionna, tasking them to \textbf{replicate Sionna radio maps} (hence, ray-tracing input to ray-tracing output). We thus verify whether the DL models can learn to generate radio maps (similarly to Sionna), and whether they bring any computational efficiency benefits.  
  
 \textbf{\texttt{Ray-tracing to Measurement Data (R2M)}}: We train the models on ray-tracing data and evaluate their accuracy against the real-world RSRP maps (ray-tracing input to real-world RSRP maps output). This comparison, largely absent from the literature, addresses whether \ac{DL} models developed with synthetic data offer accuracy gains beyond reduced computational cost. It also establishes a baseline to quantify improvements when training directly on real-world data in the following experiment.

 \textbf{\texttt{Measurement to Measurement Data (M2M)}}: We train the DL models directly on the MNO's RSRP radio maps (real-world RSRP maps input to real-world RSRP maps output). The goal is to explore whether distilling knowledge from large-scale real-world measurement data can improve the fidelity of DL models compared to ray-tracing. Further, juxtaposing these results with those of R2M allows quantifying the improvement in the DL model accuracy, stemming from the use of measured data instead of simulated data. 

 \textbf{\texttt{Site-Specific Calibration}}: This scenario aims at exploiting the differentiable nature of Sionna to learn from real-world data. It allows us to make a direct and fair comparison of the learning capacity of these approaches, as well as their potential to acquire knowledge on-the-fly.
\vspace{-4mm}

\subsection{Differentiable Ray-Tracing Simulations}
\label{RT-simulations}

For an extensive evaluation, we generate a synthetic dataset using Sionna \cite{Sionna_General}. In particular, we utilize the XML files from Section~\ref{sec:Dig_Prop_Env} to reconstruct the 3D digital propagation environment and, combined with the digital network representation of Section~\ref{sec:Digital_Net_Rep}, we create a digital twin of the MNO's RAN.
In total, we conduct 11,548 ray-tracing simulations, generating an equal number of radio maps. 
All our experiments rely on Sionna 1.1.0, and for all simulations, we launch $5\cdot 10^7$ rays, tracing up to 7 reflections. To quantify the speed improvement, we also run the same simulations using Sionna 0.19. The average simulation time on a GPU is only 0.1 seconds, whilst with the previous Sionna version, where our experiments could not fit on a GPU, the corresponding number on a server with 128 CPUs is  86.02 seconds, \ie a remarkable two orders of magnitude reduction. That buttresses the claim that the computational cost does not lie in the method \textit{per se}, but in its implementation.


\subsection{Deep Learning Baselines} \label{sec:DL_Models}

To ensure the generalizability of the DL models across different scenes, it is necessary to design suitable input features that will ensure that the models can comprehend the factors that affect signal attenuation. The descriptions from Section~\ref{sec:DL-tensor-representation} are precisely tailored to achieve this goal. Indeed, the city layout map informs the model of obstructions and their degree of influence (height), the land-use raster shows how different types of terrain affect radio propagation, whilst the road network indicates pathways for the propagating signal. Moreover, as described in Section~\ref{sec:DL-tensor-representation}, $\mathbf{L}$  allows a DL model to comprehend the impact of the antenna radiation pattern and frequency on radio wave propagation.  

Merging the four grids yields an input tensor $\mathbf{X} \in \mathbb{R}^{W \times H \times 4}$, whose elements assume a one-to-one mapping to the ground truth radio maps generated in Section~\ref{sec:Radio_Maps}. Consequently, the task of the DL-based model would be to leverage this tensor to produce high-fidelity radio maps, either through end-to-end mapping or conditional generative inference.  


To assess the potential of DL to generate high-fidelity radio maps and benchmark it with Sionna, we consider  the following representative models:

  \textbf{ \texttt{U-Net}:}  We implement a  U-Net architecture~\cite{UNet},   with a kernel size of 5, an initial feature size of 64, and a depth of 5 encoder–decoder stages. The encoder consists of sequential convolutional blocks, with two convolutional layers, a batch normalization and leaky ReLU activation,  followed by 2×2 max pooling, doubling features at each level up to 1024. 
   A bottleneck connects to a decoder that performs bilinear upsampling, convolution, and uses skip connections.  Each decoder stage mirrors the encoder structure, and a final $1\times1$ convolution outputs the RSRP radio map.

\begin{table*}[t!]
\centering
\small
\resizebox{0.995\textwidth}{!} {
\begin{tabular}{l l | c c | c | c | c | c | c }
\textbf{Case} & \textbf{Metric} & \multicolumn{2}{|c|}{\textbf{Sionna}} &  \textbf{ \texttt{U-Net}} & \textbf{ \texttt{MaxViT}}  & \textbf{ \texttt{GAN MaxViT}} & \textbf{ \texttt{MPNN}} & \textbf{ \texttt{D-MPNN}} \\
\cline{3-4}
& & \multicolumn{1}{|c}{\textbf{w/ Coverage}} & \textbf{w/o Coverage}    & & & & & \\
\hline

\multirow{5}{*}{\makecell[l]{\textbf{ \texttt{R2R,}}\\\textit{Sec.\ref{Sec:R2R}}}}
& RMSE & -- & -- &  $ 11.88 \pm 0.37$ & \cellcolor{blue!05} $ \mathbf{10.74 \pm 0.35}$ &  $11.06 \pm 0.31$ & $14.88 \pm 0.53$ & $15.77 \pm 1.01$ \\
& MAE & -- & -- &   $8.21 \pm 0.41$ & $7.64 \pm 0.32$ & \cellcolor{blue!05} $\mathbf{7.03 \pm 0.55}$ & $11.26 \pm 0.64$ & $12.92 \pm 0.98$ \\
& SSIM & -- & -- &  $0.49 \pm 0.04$ & \cellcolor{blue!05} $\mathbf{0.50 \pm 0.03}$ & \cellcolor{blue!05} $\mathbf{0.50 \pm 0.03}$ & $0.48 \pm 0.05$ & $0.43\pm 0.05$ \\
& PCC    & -- & -- &   $0.74 \pm 0.02$ & \cellcolor{blue!05} $\mathbf{0.77 \pm 0.02}$ & $0.76 \pm 0.02$ &  $0.58 \pm 0.03$  & $0.58 \pm 0.03$ \\
& SMAPE & -- & -- &   $0.03 \pm 0.002$ &  $0.03 \pm 0.001$ & $0.03 \pm 0.002$ &  $0.04 \pm 0.003$  & $0.04 \pm 0.003$ \\

\hline
\hline

\multirow{5}{*}{\makecell[l]{\textbf{ \texttt{R2M}}\\\textit{Sec.\ref{Sec:R2M}}}}
& RMSE & $20.12 \pm 2.41$ & \cellcolor{blue!05} $ \mathbf{ 15.77 \pm 1.01}$   & $17.57 \pm 1.41$ & $ 16.83 \pm 2.19$ &  $17.57 \pm 0.28$ & $18.86 \pm 1.35$ &  $18.26 \pm 0.66$ \\
& MAE & $16.52 \pm 2.22$ & \cellcolor{blue!05} $ \mathbf{12.92 \pm 0.98}$   & $14.68 \pm 1.29$ & $14.02 \pm 2.12$ & $14.49 \pm 2.25$ & $15.08 \pm 1.33$ &  $15.44 \pm 0.39$ \\ 
& PCC  & $0.34 \pm 0.02$  & $0.35 \pm 0.02$   & $0.44 \pm 1.69$  &  $0.42 \pm 1.58$ & $0.41 \pm 2.48$ & $0.26 \pm 0.04$ & $0.34 \pm 0.02$ \\
& SMAPE & $0.08 \pm 1.01$ & $0.06 \pm 0.005$   & $0.07 \pm 0.007$   & $0.07 \pm 0.01$  & $0.07 \pm 0.01$  & $0.09 \pm 0.01$ & $0.08 \pm 0.002$ \\

\hline
\hline

\multirow{5}{*}{\makecell[l]{\textbf{ \texttt{M2M}}\\\textit{Sec.\ref{Sec:M2M}}}} 
& RMSE & $20.12 \pm 2.41$ & $ 15.77 \pm 1.01 $ &   $10.30 \pm 0.53$ & $10.03 \pm 0.44$ & \cellcolor{blue!05} $ \mathbf{9.85 \pm 0.88}$ & $11.72 \pm 0.78$ & $11.20 \pm 0.89$\\
& MAE & $16.52 \pm 2.22$ & $12.92 \pm 0.98$ &   $8.31 \pm 0.48$ & $8.01 \pm 0.38$ &\cellcolor{blue!05} $ \mathbf{7.76 \pm 0.66}$  & $9.64 \pm 0.70$ & $9.19 \pm 0.83$ \\
& $R^2$  & $34.29 \pm 2.15$  & $0.35 \pm 0.02$ &   $0.51 \pm 0.02 $ & $0.51 \pm 0.02 $ &  \cellcolor{blue!05} $ \mathbf{0.52 \pm 0.05 }$ & $0.32 \pm 0.03 $ & $0.32\pm 0.07$ \\
& SMAPE & $8.29 \pm 1.01$ & $0.06 \pm 0.005$ &  $0.04 \pm 0.003 $ & $ 0.04 \pm 0.002$ & $0.04 \pm 0.004$& $0.04\pm 0.005$  & $0.05\pm 0.005$ \\

\hline
\hline 

\multirow{3}{*}{\textbf{Complexity}} 
& {Size (M)} & -- & -- &  216.526  & 221.838 & 222.535& 0.07 & 2.43 \\

& {FLOPs (G)} & -- & -- &   645.972   & 271.723 & 281.655 & 84.192  & 185 \\
& {Runtime (ms)} & $ 115.95 \pm 129.40$  &  $ 115.95 \pm 129.40$ &   $3.38\pm 4.12$ & $49.83 \pm 5.36$ & $49.83 \pm 5.36$ & \cellcolor{blue!05} $ \mathbf{1.11 \pm 2.31}$ & $7.17\pm 6.04$\\

\end{tabular}
}
\vspace*{2pt}
\caption{Comparison of radio propagation models under different learning conditions: \textbf{Ray-tracing to \mbox{Ray-tracing}~(\texttt{R2R})}, \textbf{Ray-tracing to Measurement Data (\texttt{R2M})}, and \textbf{Measurement to Measurement Data (\texttt{M2M})}. }
\label{tab:generalization_R2R_R2M_M2M}
\vspace*{-2pt}
\end{table*}
 
\textbf{ \texttt{MaxVit}:} We employ a MaxViT-Large module as the encoder backbone, combining MBConv blocks \cite{MBconv} with multi-axis self-attention  to capture both global and local features. Across all attention layers, each block has attention heads of size 32, and each MaxViT block comprises an MBConv inverted bottleneck with an expansion rate of 4, followed by a squeeze-and-excitation module \cite{Squeeze_and_Excitation} with a shrinkage ratio of 0.2. The hierarchical design reduces resolution and increases channels, producing multi-scale feature maps that are fused through U-Net-style skip connections and are converted back to the RSRP map in the decoding stage.
    
\textbf{  \texttt{GAN MaxViT}:} We integrate the  \textbf{\texttt{MaxVit}} model presented earlier as the generator of a conditional GAN (cGAN) \cite{GAN_Paper}. For the discriminator, we employ the PatchGan model, which outputs a tensor indicating whether patches of $70\times70$ pixels extracted from the 
cGAN generator are realistic or not. The discriminator minimizes binary cross-entropy (BCE) loss over real and fake pairs, while the generator is optimized using a combination of adversarial BCE loss and L1 reconstruction loss.

\textbf{  \texttt{Message-Passing Neural Network (MPNN):} } We use the input tensor representation (Section~\ref{sec:DL-tensor-representation}) to extract a graph and infer the RSRP radio maps through message-passing between the graph nodes \cite{Message_Passing}. Specifically, each grid element in the tensor is considered as a graph node, assigned with the commensurate features across grids. Then, four types of edges are defined by connecting: ($i$) the Tx node to all nodes, ($ii$) first-hop neighbors, 
   ($iii$)  second-hop road neighbors, 
   ($iv$) 
   all nodes within the same building. Each edge assumes three features: the physical distance between the nodes and the relative Cartesian displacement. The message-passing operation is  performed via a three-layer MLP (32, 64, 128 units) on the concatenated node-edge features. 
   The aggregated message embeddings are passed along with the current node representation to an update function realized via a two-layer MLP (128 and 256 units). A final output layer maps the updated representation to a scalar value per node  representing the inferred RSRP. 

\textbf{  \texttt{D-MPNN:}} Instead of a single graph with multiple types of connections, we consider four distinct graphs, each one processed by a dedicated MPNN. All graphs share identical node features,  but each graph has only one of the four edge types described earlier. The dedicated MPNNs applies message-passing via a three-layer MLPs (32, 64, and 128 units) followed by an update function implemented via a two-layer MLP (128 and 256 units). The four node embeddings produced by the dedicated MPNNs are then concatenated and forwarded to an output two-layer MLP module with 512 and 256 neurons that performs a progressive dimensionality reduction to a single scalar output per node.

All models are trained using the AdamW optimizer with a learning rate of $ 10^{-3}$ and a weight decay of $5\cdot 10^{-2}$, reducing the learning rate by a factor of 0.7 when the validation loss failed to improve for 10 consecutive epochs. The batch size for \textbf{\texttt{U-Net}} is 4, for \textbf{\texttt{MaxVit}} and the \textbf{\texttt{GAN}} 2, while for the GNNs it is 1. Apart from the GAN, the loss function to minimize for the rest models is the mean squared error (MSE). All modes are deployed in a 24 GB NVIDIA RTX A5000.
 
\subsection{Learning from Ray Tracing}   \label{Sec:Learning_Synthetic_Data}


We use the Sionna-generated radio maps to train the DL models of Section~\ref{sec:DL_Models}. To ensure the robustness and scalability of the training process, we perform a 5-fold cross-validation. For each cross-validation fold, we select a subset of the 13 Areas for training and use the remaining Areas for testing, ensuring that the test set covers at least $20\%$ of the total samples. During training, we evaluate the model from each fold on $10\%$ of the training data that is kept out as a validation set, and the best model is picked. 
Note that Areas are assigned to folds such that training and test sets are disjoint. 
 While some Areas may appear in the test set more than once, we choose the assignment so as to minimize repetition and maintain a balanced evaluation. This process ensures that the DL models can act as standalone solvers, suitable for MNOs, and directly generate radio maps for new Areas with characteristics different from those seen during training. To evaluate the DL model performance, we employ a collection of metrics, including the root mean square error (RMSE), mean absolute error (MAE), the structural similarity index measure (SSIM), the Pearson correlation coefficient (PCC), and the symmetric mean absolute percentage error (SMAPE). The corresponding expression for each metric can be found in Appendix~\ref{sec:Appendix_Metrics_R2R}.
 
\subsubsection{\textbf{Ray-tracing to Ray-tracing (\texttt{R2R}})} \label{Sec:R2R} 
 
Table~\ref{tab:generalization_R2R_R2M_M2M} shows the performance of the best models picked from the cross validation with respect to the synthetic ray-tracing data. The results were obtained by evaluating the models over the test set of the 
corresponding fold. Overall, both convolutional (\textbf{\texttt{U-Net}}) and transformer-based (\textbf{\texttt{MaxViT}}, \textbf{\texttt{GAN MaxViT}}) models outperform graph-based approaches such as \textbf{\texttt{MPNN}} and \textbf{\texttt{D-MPNN}}, showing weaker performance in this task, suggesting that spatially structured models are better suited for this surrogate modeling scenario.  Specifically, transformer-based architectures, \ie \textbf{\texttt{MaxViT}} and \textbf{\texttt{GAN MaxViT}}, attain the lowest errors, both in end-to-end and GAN-based configurations, yielding an RMSE and MAE at around 10.74 and 7.03 dB, respectively. These results demonstrate the potential of the proposed models to accurately reproduce radio maps, which can be consequently leveraged to fuse knowledge from measurements.


\subsubsection{\textbf{Ray-tracing to Measurements (\texttt{R2M}})}\label{Sec:R2M} We now explore how the predictions of the DL models compare with the real-world RSRP radio maps of Section~\ref{sec:Radio_Maps}. In addition, we also explore the \textit{out-of-the-shelf} accuracy of Sionna for the same scenes without tuning its parameters with real-world data.  At this point, it is important to note that in dense urban environments, Sionna outputs a considerable \mbox{number} of \mbox{no-coverage} points, \ie points not reached by rays. 
These null points are filled with the smallest RSRP value encountered in our dataset, \ie -140 dBm, which coincides with receiver sensitivity limits. For the maps in each cross-validation set, the average coverage percentages ranges from $79.84\%$ to $91.62\%$. As this is an inherent limitation of the method, to ensure a fair and comprehensive evaluation, we provide the performance breakdown for both including and excluding the no-coverage points in the metric computation.

\begin{table*}[h!]
\centering
\small
\resizebox{\textwidth}{!} {
\begin{tabular}{l | cc | cc | cc | cc | c | c }
\textbf{Area} & 
\multicolumn{2}{c|}{\textbf{Sionna}} & 
\multicolumn{2}{c|}{\textbf{Sionna-A}} & 
\multicolumn{2}{c|}{\textbf{Sionna-AM}} & 
\multicolumn{2}{c|}{\textbf{Sionna-AMv}} & 
 \textbf{\texttt{GAN MaxViT-M2M}} &  
 \textbf{\texttt{GAN MaxViT-C}} \\
\cline{2-11} 
&  Cov. & w/o Cov.
&  Cov. &  w/o Cov.
&  Cov. &  w/o Cov. 
&  Cov. &  w/o Cov. 
& &  \\
\hline

Area 1  & $13.19 \pm 3.50$ & $15.33 \pm 4.88$ & $9.43 \pm 1.94$ & $15.74 \pm 6.01$ & $8.62 \pm 2.69$ & $18.35 \pm 6.22$ & $7.63 \pm 1.44$ & $14.31 \pm 6.04$ & $7.53 \pm 1.20$ & \cellcolor{blue!5} $\mathbf{4.70 \pm 0.41}$   \\ 
Area 2  & $12.22 \pm 2.93$ & $14.25 \pm 4.79$ & $10.22 \pm 1.71$ & $17.19 \pm 6.10$ & $9.44 \pm 2.81$ & $19.20 \pm 5.63$ & $7.67 \pm 1.16$ & $15.38 \pm 6.35$ & $7.91 \pm 1.79$ & \cellcolor{blue!5} $\mathbf{4.69 \pm 0.43}$   \\ 
Area 3  & $14.17 \pm 5.98$ & $17.73 \pm 7.53$ & $11.05 \pm 2.13$ & $21.99 \pm 6.18$ & $9.82 \pm 3.46$ & $23.22 \pm 5.10$ & $7.66 \pm 1.05$ & $19.00 \pm 5.95$ & $6.94 \pm 1.67$ & \cellcolor{blue!5} $\mathbf{4.53 \pm 0.49}$   \\ 
Area 4  & $12.48 \pm 3.18$ & $16.09 \pm 5.78$ & $11.33 \pm 2.15$ & $20.71 \pm 6.90$ & $11.07 \pm 3.63$ & $22.39 \pm 5.32$ & $7.78 \pm 1.12$ & $18.64 \pm 7.78$ & $7.76 \pm 1.36$ & \cellcolor{blue!5} $\mathbf{4.88 \pm 0.49}$   \\ 
Area 5  & $11.70 \pm 2.87$ & $12.73 \pm 3.52$ & $10.36 \pm 1.68$ & $15.97 \pm 4.18$ & $9.27 \pm 2.43$ & $18.20 \pm 3.75$ & $7.70 \pm 0.90$ & $13.33 \pm 3.72$ & $8.15 \pm 1.56$ & \cellcolor{blue!5} $\mathbf{4.88 \pm 0.46}$    \\ 
Area 6  & $17.97 \pm 8.29$ & $23.25 \pm 11.77$ & $11.16 \pm 2.94$ & $26.24 \pm 11.72$ & $11.14 \pm 3.26$ & $28.23 \pm 11.71$ & $7.65 \pm 1.99$ & $26.53 \pm 16.38$ & $9.85 \pm 2.39$ & \cellcolor{blue!5} $\mathbf{4.95 \pm 0.51}$   \\ 
Area 7  & $14.46 \pm 7.04$ & $19.43 \pm 12.12$ & $11.42 \pm 2.60$ & $19.45 \pm 8.86$ & $10.01 \pm 3.34$ & $21.89 \pm 8.08$ & $8.10 \pm 1.32$ & $18.55 \pm 10.03$ & $7.43 \pm 1.36$ & \cellcolor{blue!5} $\mathbf{4.65 \pm 0.53}$   \\ 
Area 8  & $12.29 \pm 3.12$ & $14.97 \pm 7.05$ & $10.03 \pm 2.24$ & $15.95 \pm 6.31$ & $8.87 \pm 2.72$ & $19.15 \pm 6.96$ & $7.65 \pm 0.97$ & $ 14.13 \pm 6.03$ & $7.64 \pm 1.47$ & \cellcolor{blue!5} $\mathbf{4.92 \pm 0.42}$   \\ 
Area 9  & $12.64 \pm 3.34$ & $14.21 \pm 4.22$ & $10.03 \pm 2.06$ & $16.79 \pm 5.44$ & $8.29 \pm 2.49$ & $19.65 \pm 4.89$ & $7.64 \pm 1.12$ & $14.53 \pm 4.98$ & $7.45 \pm 1.25$ & \cellcolor{blue!5} $\mathbf{4.70 \pm 0.44}$   \\ 
Area 10 & $13.55 \pm 5.02$ & $18.26 \pm 9.59$ & $8.12 \pm 0.00$ & $20.72 \pm 7.53$ & $9.65 \pm 3.89$ & $24.14 \pm 8.97$ & $8.03 \pm 1.40$ & $16.40 \pm 6.47$ & $8.87 \pm 1.97$ & \cellcolor{blue!5} $\mathbf{4.93 \pm 0.56}$   \\ 
Area 11 & $12.14 \pm 3.25$ & $14.56 \pm 5.09$ & $10.26 \pm 2.29$ & $18.27 \pm 6.34$ & $9.33 \pm 3.10$ & $20.50 \pm 6.30$ & $7.84 \pm 1.44$ & $19.39 \pm 9.59$ & $7.79 \pm 1.51$ & \cellcolor{blue!5} $\mathbf{4.93 \pm 0.56}$   \\ 
Area 12 & $11.53 \pm 2.86$ & $13.38 \pm 4.37$ & $10.58 \pm 2.07$ & $18.13 \pm 5.83$ & $9.59 \pm 3.15$ & $20.44 \pm 5.50$ & $7.82 \pm 1.32$ & $16.01 \pm 5.81$ & $7.57 \pm 1.60$ & \cellcolor{blue!5} $\mathbf{4.84 \pm 0.73}$   \\ 
Area 13 & $12.32 \pm 3.32$ & $16.67 \pm 8.42$ & $11.21 \pm 2.22$ & $20.72 \pm 7.53$ & $10.10 \pm 3.05$ & $22.18 \pm 6.67$ & $7.84 \pm 1.44$ & $19.39 \pm 9.59$ & $8.06 \pm 1.50$ & \cellcolor{blue!5} $\mathbf{5.12 \pm 0.38}$   \\ 

\hline
\end{tabular}
}
\vspace*{2pt}
\caption{Site-specific calibration performance breakdown table for various models.}
\label{tab:Calibration}
\vspace*{-2mm}
\end{table*}

As shown in Table~\ref{tab:generalization_R2R_R2M_M2M}, Sionna attains the smallest errors by excluding no-coverage points. Specifically, it yields an RMSE of $15.77 \pm 1.01$ and a MAE of $12.92 \pm 0.98$, outperforming both \textbf{\texttt{U-Net}} and \textbf{\texttt{MaxViT}} variants, which show comparable but slightly higher errors. However, the evaluation without coverage masking inflates the reported errors, as no-coverage areas are explicitly set to -140 dBm, resulting in larger deviations. Importantly,  the results corroborate the intuition that models developed with synthetic data will perform at the same level when compared to real-world data. At the same time, the measured runtimes (Table~\ref{tab:generalization_R2R_R2M_M2M}) indicate that Sionna can generate radio maps at speeds comparable to the best DL model, with the previously reported orders-of-magnitude runtime gap now drastically reduced and emulation times remaining well under one second. 


\begin{tcolorbox}[maintakeaway]
 \textit{\textbf{Key insights}}: DL models can learn to replicate synthetic radio maps from high-performance propagation solvers. However, their accuracy levels with respect to real-world data remain comparable, whilst the unprecedented speed of Sionna has sharply narrowed or eliminated their computational efficiency benefits. That calls into question the motivation for pursuing DL-based models.
\end{tcolorbox}


\subsection{Learning from real measurements} \label{Sec:M2M}

Although the computational speed advantage may be diminished, the primary strength of DL models lies in their ability to leverage real-world data and achieve higher accuracy than conventional models. To explore this potential, instead of training the DL models with synthetic data,  we now use directly the ground truth radio maps of Section~\ref{sec:Radio_Maps}. All the model parameters and training scheme remain the same, \ie we conduct a 5-fold cross-validation over the same set of areas with the same model hyperparameters. In this case, as per Appendix~\ref{sec:Appendix_Metrics_R2M}, the loss function and evaluation metrics are applied only for grid points for which there are measurement data.

\subsubsection{\textbf{Measurement to Measurement data (\texttt{M2M}})}
 The results from this set of experiments are presented in Table~\ref{tab:generalization_R2R_R2M_M2M}. As expected, training DL models with real-world data, instead of synthetic, substantially increases their fidelity and reduces their error levels. Indeed, the reported error metrics are improved by approximately $40\%$, exhibiting roughly a 7dB improvement in the RMSE and MAE, compared to when trained with ray-tracing data. The training set-up ensures that this improvement occurs in settings that the model has not been exposed to, \ie unknown areas and antennas,  thus attesting to their generalizability and their potential to be used as standalone solvers.  
 
 Importantly, the inferred RSRP estimates in the $\textbf{\texttt{M2M}}$ case are more reliable than these of differentiable ray-tracing; DL models are $35\%$ more accurate, which translates into a $\approx$5 dB error metric difference.  Yet, the fundamental open question is: \textit{Can the unprecedented speed of Sionna, combined with its differentiable nature, match or exceed this accuracy improvement and, consequently, depose DL-based models?} The next subsection answers this key question.


\begin{tcolorbox}[maintakeaway]
\textit{\textbf{Key insights}}: While the efficiency gap has been reduced, DL models still preserve a leading position due to their potential to learn from real-world data. That enables them to improve their fidelity and outperform differentiable ray-tracing in terms of accuracy. Hence, DL-based radio map generation can yield  significant improvements, provided they are developed with real-world data.
\end{tcolorbox}

\subsection{Site-Specific Calibration}

In contrast with DL-based models, the differentiable nature of Sionna exhibits the potential to learn from real-world data in a site-specific context by tuning the antenna radiation pattern parameters and the electromagnetic properties of the materials within the propagation environment. Hence, 
unlike the  \textbf{\texttt{M2M}} case, where the models adapted their weights effectively and demonstrated generalizability for unknown scenes not used during training, in Sionna, the antenna and material properties for one scene are not transferable to another scene, forcing the calibration process to be performed from scratch. While that could potentially be a limitation, the ``lighting-fast'' nature of Sionna could enable learning on-the-fly from data generated in the MNO's network, thus mitigating the generalizability limitation. 

In this section, we thoroughly assess the potential of Sionna and benchmark it with that of DL-based models. To this end, for each scene we randomly select $70\%$ of the available measurements to be used in the calibration process, whilst the remaining $30\%$ is kept to evaluate its effectiveness. Note that we allocate $30\%$ of the measurements to the validation set, rather than the  $20\%$ used in previous sections. This is to provide a more reliable assessment of model generalization while retaining sufficient data for effective training. To have an extensive, commensurate and fair comparison between differentiable ray-tracing and DL models, we consider the following case for the latter: take the best pre-trained model of Section~\ref{Sec:M2M}, \ie \textbf{\texttt{GAN MaxViT}}, and task it to directly infer the  RSRP for the validation set. We refer to that model as \textbf{\texttt{GAN MaxViT-M2M}}. 
Second, we take the pre-trained  \textbf{\texttt{GAN MaxViT}} instance, and we retrain it separately per scene to learn to predict the RSRP for the scene measurements reserved for calibrating the differentiable ray-tracer in that scene. This process is directly analogous to Sionna's calibration concept, \ie a full-fledged propagation model assimilates knowledge from a single scene by adapting its parameters, without, though, necessitating generalizability; we refer to it as \textbf{\texttt{GAN MaxViT-C}}.

\begin{figure}
\subfloat{\includegraphics[width=0.75\columnwidth]{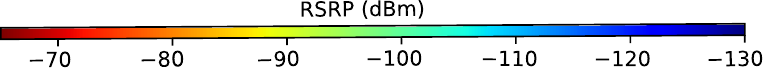}}
\vspace{1pt}

\centering
\subfloat{\includegraphics[height=0.2\columnwidth]{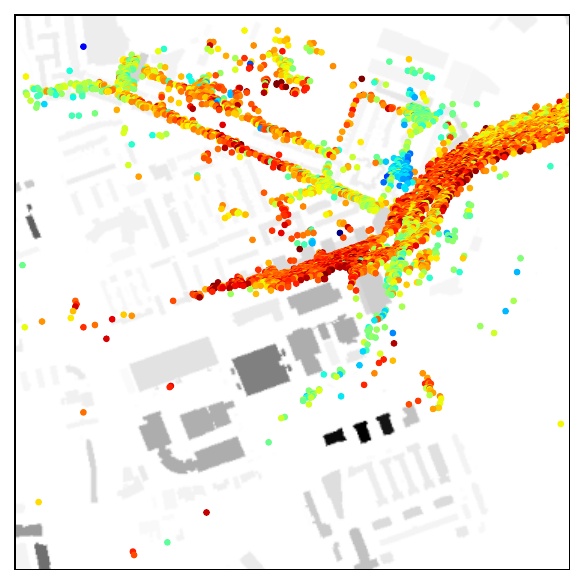}}
\subfloat{\includegraphics[height=0.2\columnwidth]{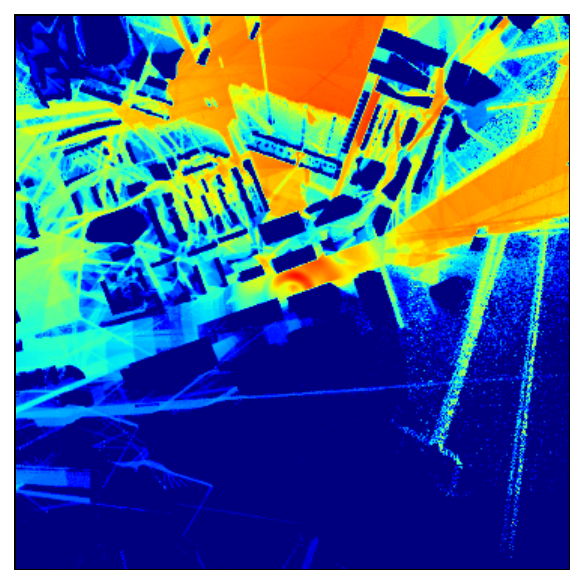}}
\subfloat{\includegraphics[height=0.2\columnwidth]{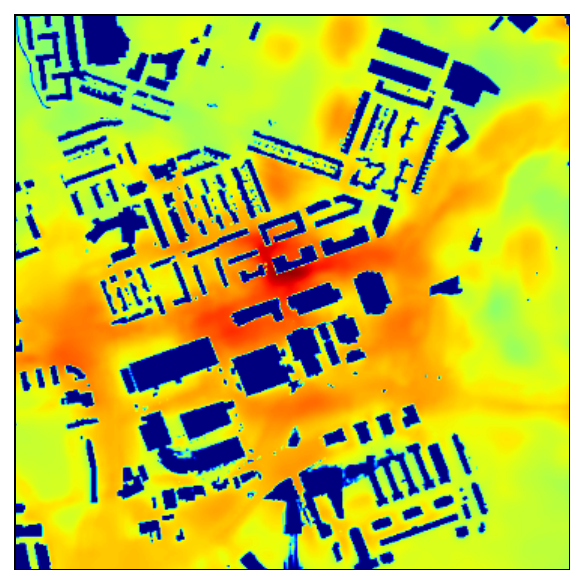}}
\subfloat{\includegraphics[height=0.2\columnwidth]{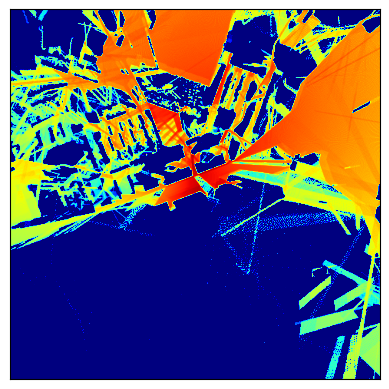}}
\subfloat{\includegraphics[height=0.2\columnwidth]{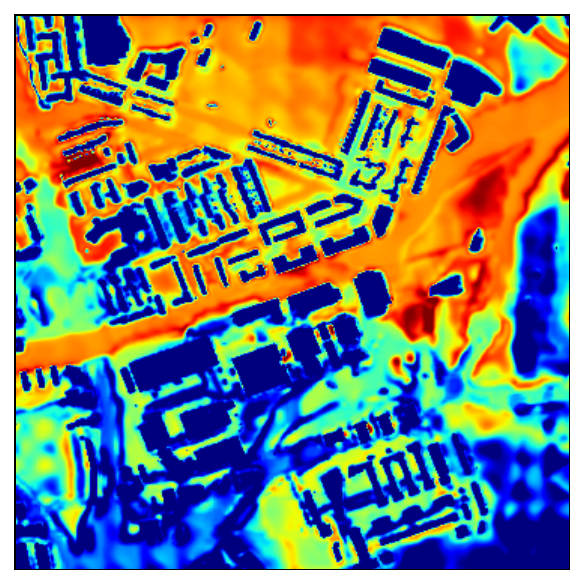}}

\subfloat{\includegraphics[height=0.2\columnwidth]{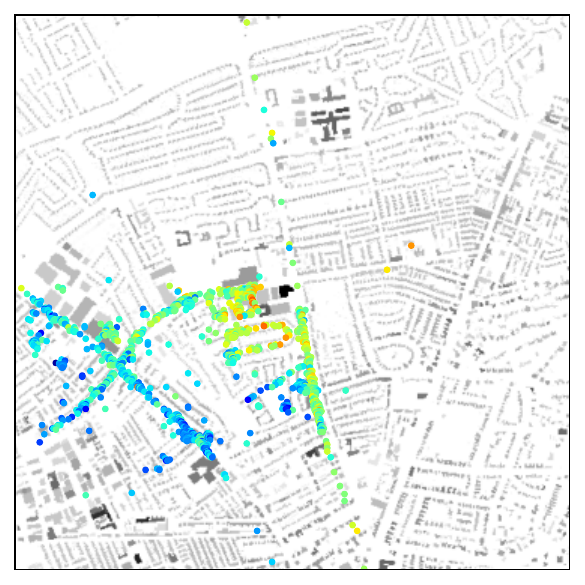}}
\subfloat{\includegraphics[height=0.2\columnwidth]{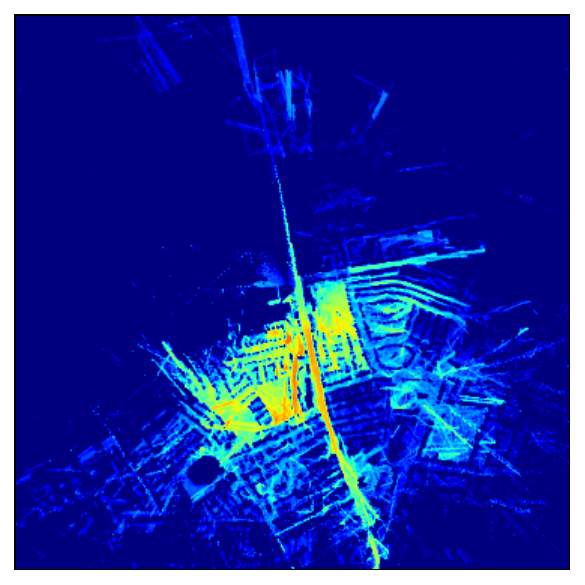}}
\subfloat{\includegraphics[height=0.2\columnwidth]{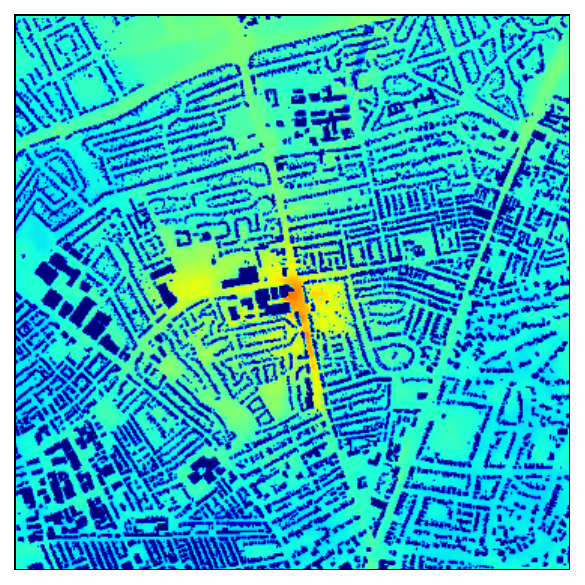}}
\subfloat{\includegraphics[height=0.2\columnwidth]{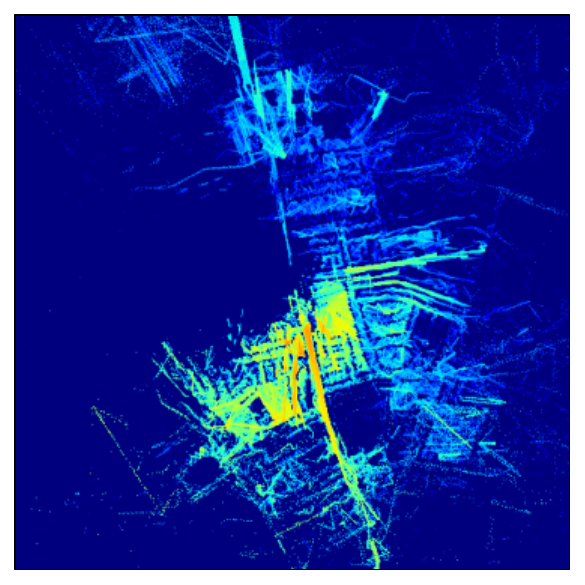}}
\subfloat{\includegraphics[height=0.2\columnwidth]{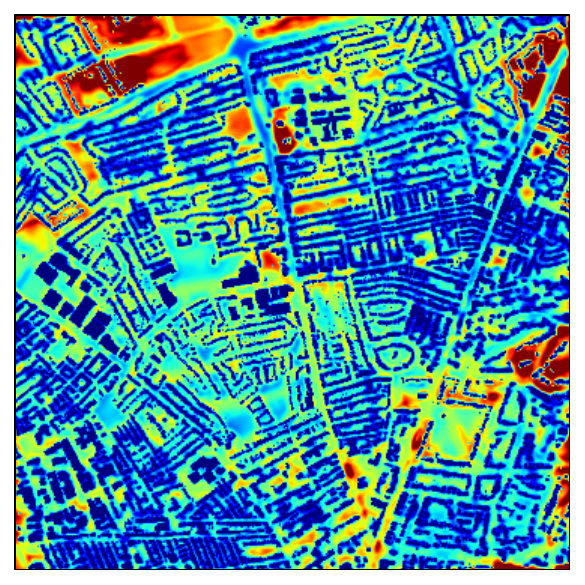}}

\captionsetup[sub]{font=scriptsize}
\setcounter{subfigure}{0}
\subfloat[Measured Data]{\includegraphics[height=0.2\columnwidth]{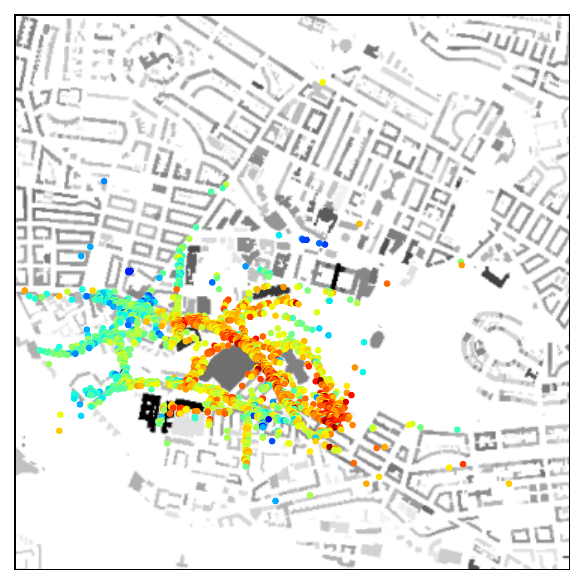}}
\subfloat[Sionna]{\includegraphics[height=0.2\columnwidth]{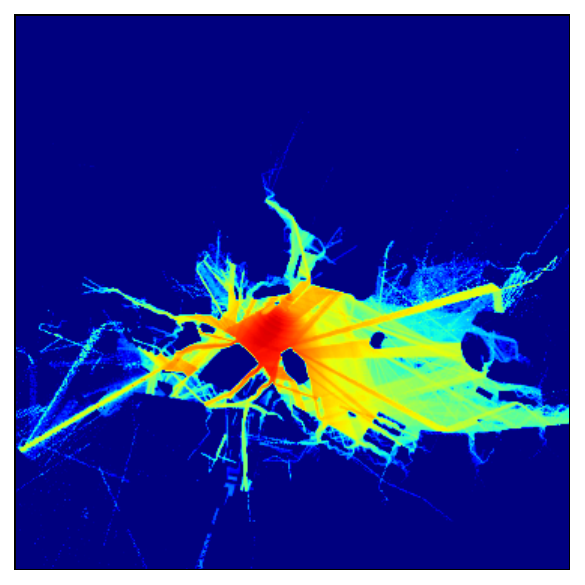}}
\subfloat[GAN MaxViT-M2M]{\includegraphics[height=0.2\columnwidth]{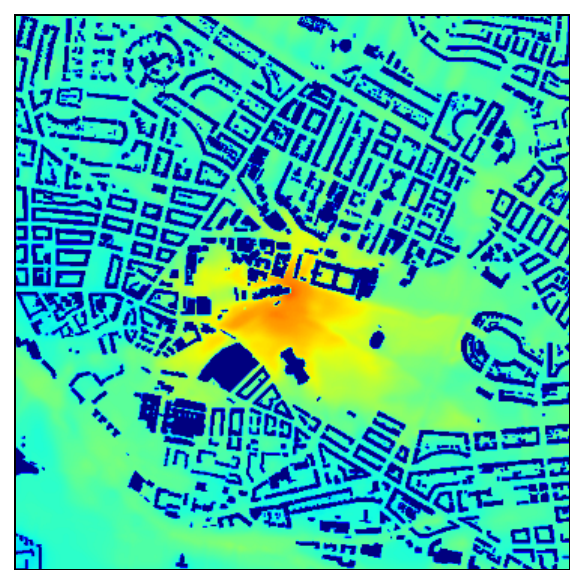}}
\subfloat[Sionna-AMv]{\includegraphics[height=0.2\columnwidth]{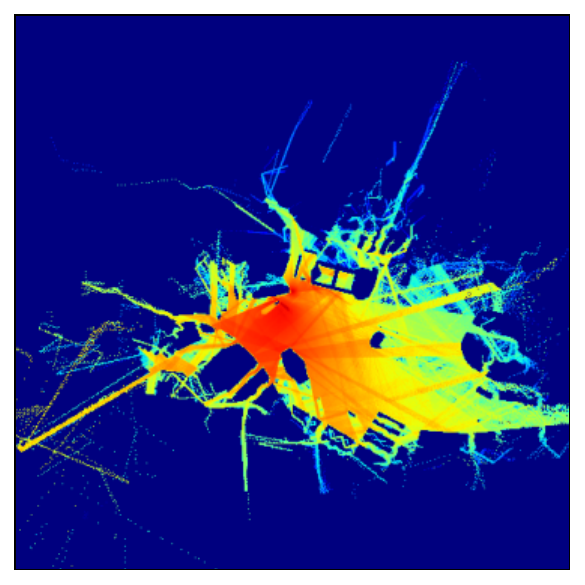}}
\subfloat[GAN MaxViT-C]{\includegraphics[height=0.2\columnwidth]{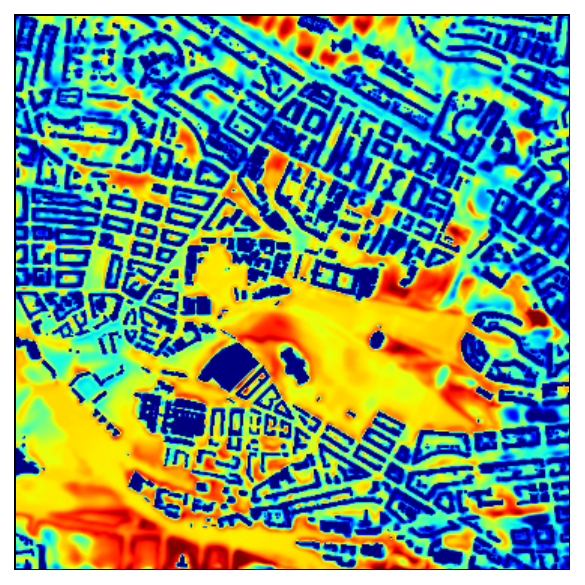}}
 
\caption{Visual comparison of generated RSRP maps before and after callibration.}
\label{fig:best_bacelli_model_prediction}
 
\end{figure}

\subsubsection{\textbf{Calibrated ray-tracing simulations}}
To make \linebreak Sionna trainable, we consider the following cases. First, we follow the official Sionna developer guidelines\footnote{https://nvlabs.github.io/sionna/rt/developer/developer.html} to design a trainable version of the \texttt{tr38901} antenna radiation pattern, setting learnable parameters for the azimuth steering angle, the horizontal and vertical half-power beamwidth, as well as the maximum element gain, which acts as an offset.\paul{AP: Cite 3GPP.} This approach introduces four learnable parameters, and we refer to it as \textbf{Sionna-A}. Then, according to developer guidelines, we consider the case that, along with antenna parameters, we jointly calibrate all objects with the same material, assigning two trainable parameters (permittivity and conductivity) for each material type. For a scene with $M$ distinct materials, that results in $2M +4$ optimization parameters, and we typically have from 3 to 7 materials per scene; this case is denoted as \textbf{Sionna-AM}.  We note that we also experimented with trainable scattering and polarization discrimination coefficients, but we observed a deterioration in calibration quality, thus we do not consider that case.

Finally, instead of optimizing independent scalar variables via automatic differentiation, we consider a vectorized calibration parameter formulation~\cite{SIonna_Learning}. Each scalar variable is parametrized as the output of a bounded function of a trainable  \textit{read-out vector}, $v \in  \mathbb{R}^{L_v}$, projecting a learned \textit{embedding vector}, $w \in   \mathbb{R}^{L_w}$,  onto a scalar values via a dot product. Consequently, each optimization parameter, $p$, has its dedicated \textit{embedding vector} vector $w_p$, whilst we consider two shared \textit{read-out vectors}, one for the material parameters and one for the antenna parameters. According to \cite{SIonna_Learning}, we set $L_v =5$ and $L_w = 5$ for our experiments. We refer to this formulation as \textbf{Sionna-AMv}, and more details about it can be found in Appendix~\ref{sec:Appendix_Sionna_vectorized}.

At this point, it is paramount to highlight in advance several limitations we encountered in Sionna's calibration, which will also become evident from the results presented hereunder. First, Sionna has two operation modes configured by the parameter $\texttt{loop mode}$, which defines the Dr.Jit mode used to evaluate the loop that implements the solver. The first mode, \texttt{symbolic} is the default and fastest one but does not support automatic differentiation. Hence, calibration entails using the second mode, \texttt{evaluated}, which severely affects the remarkable simulation speed documented in Sections~\ref{Sec:R2R} and~\ref{Sec:M2M}. Second, Sionna by default merges all the shapes that share the same material into a single shape for increased efficiency. There is the option to fully or partially skip merging, and consequently considering the electromagnetic material properties of each building as distinct parameters. However, we observed that this leads to prohibitive simulation times, exceeding 15 minutes for a single simulation per scene. Third, we found that backpropagation using Dr.Jit is substantially more time-consuming compared to that used in common DL models development. Specifically, this step can take roughly one second when shapes are merged (comparable to the ray-tracing simulation time, thus adding a non-negligible overhead), but can exceed ten minutes when shapes are not merged. Finally, as mentioned in Section~\ref{Sec:R2M} and observed in Figure~\ref{fig:best_bacelli_model_prediction}b, Sionna yields a  large number of no-coverage points, which are set to a threshold value (-140 dBm). During the calibration process, these points are masked out, \ie not considered in the loss function evaluation.

\subsubsection{\textbf{Calibrated ray-tracing vs site-specific DL model}}

\begin{figure}
    \centering
    \includegraphics[width=0.95\columnwidth,trim={0 0 0 5pt},clip]{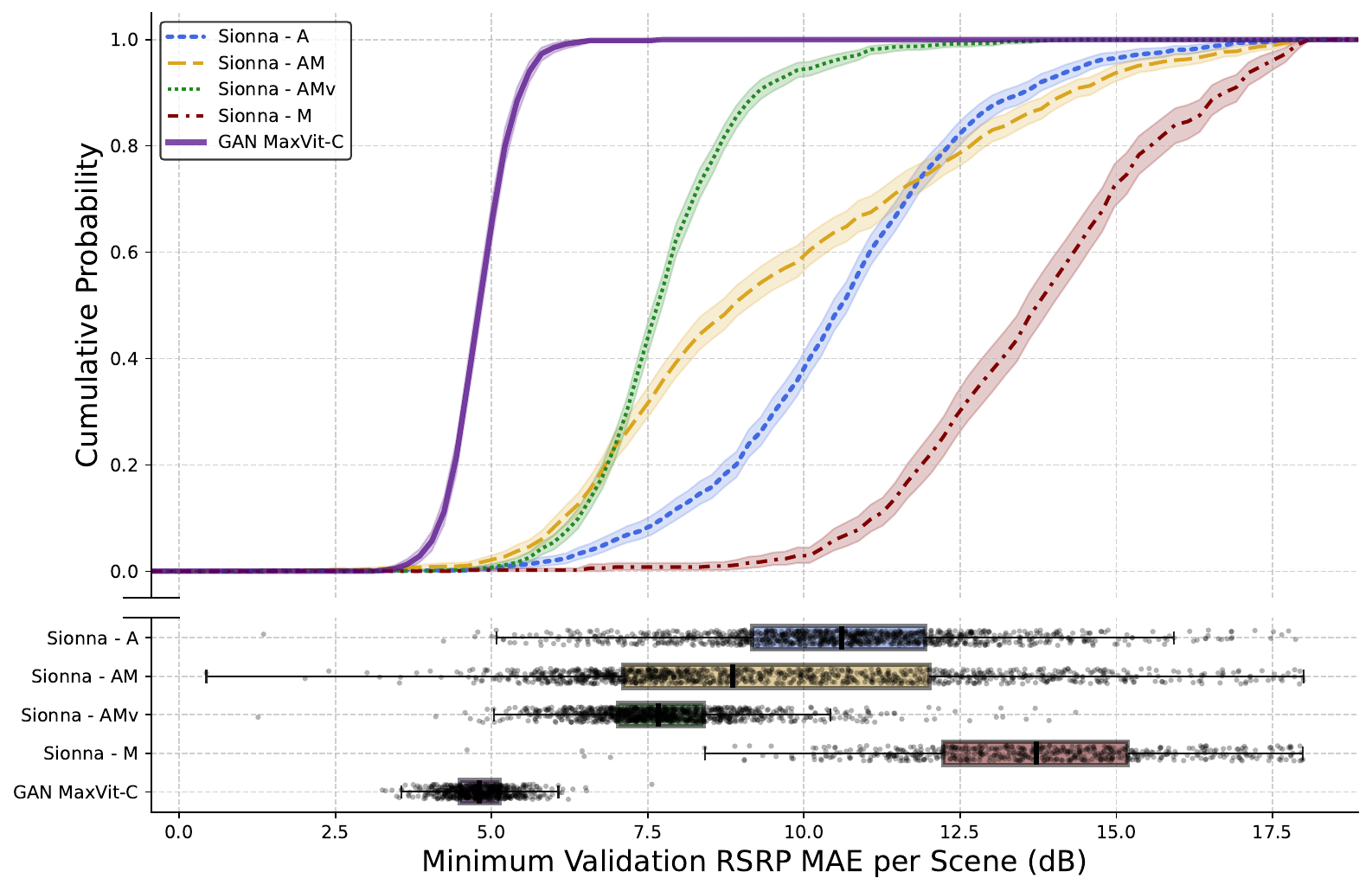}
 
    \caption{CDF and box plot of the minimum validation RSRP MAE attained per scene over the calibration.}
    \label{fig:Calibration_Result_Comparison}
 
\end{figure}

For each scene, we compute the minimum validation MAE during the calibration, achieved over all validation points in that scene, both with and without the inclusion of no-coverage points. In Table~\ref{tab:Calibration}, we report, across all scenes, for the distinct areas of our dataset, the mean and standard deviation of these per-scene minimum validation MAEs. 
Figure~\ref{fig:Calibration_Result_Comparison} complements this table by illustrating the distribution of validation MAE values via the respective CDFs and a boxplot, highlighting the overall error distribution across scenes. Quantitative results for Sionna-M are omitted from Table~\ref{tab:Calibration}, as Figure~\ref{fig:Calibration_Result_Comparison} demonstrates that it provides minimal improvement. 

Configuring only the antenna parameters or jointly optimizing antenna and material parameters reduces the error by approximately 3 dB and 3.5 dB respectively. Employing a vectorized representation further decreases the error, ensuring training stability, with all areas achieving errors close to 7.7 dB. Notably, a DL model pretrained on different areas achieves comparable errors with \textbf{Sionna-AMv} without any fine-tuning, suggesting that a properly trained DL model can effectively replace the calibration process. At the same time, applying site-specific calibration to the DL model further reduces the error across all areas to roughly 5 dB, representing a 2.7 dB improvement over the best-performing Sionna calibration variant. As illustrated in Figure~\ref{fig:best_bacelli_model_prediction}, the DL-calibrated radio maps adapt effectively their RSRP estimates and they are not affected by the pronounced no-coverage problem of Sionna. 

\begin{figure}
    \centering
    \includegraphics[width=0.95\columnwidth,trim={0 0 0 5pt},clip]{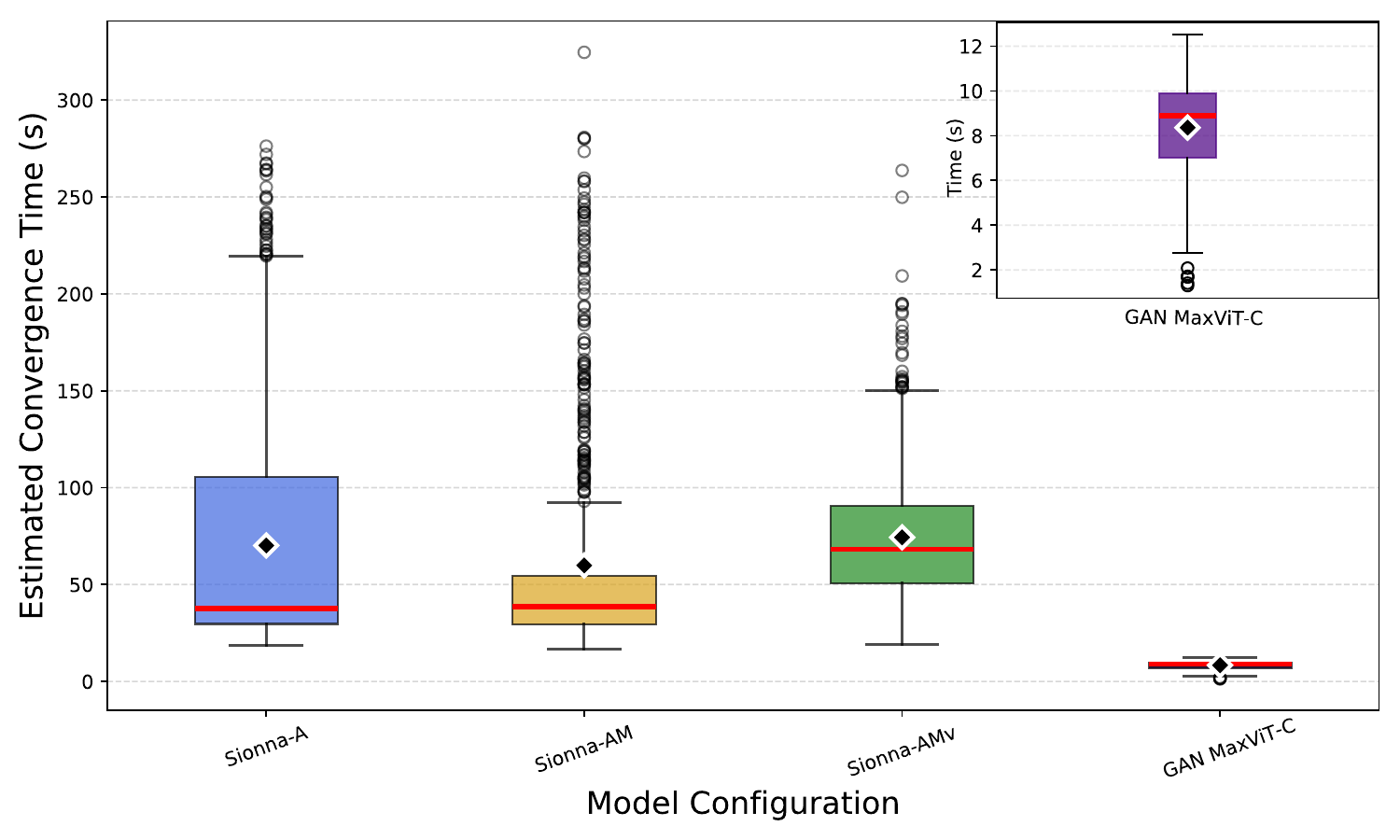}
    
    \caption{Convergence time for different calibration approaches.}
    \label{fig:Time_Calibration_Result_Comparison}
     
\end{figure}

These observations are particularly important when considering calibration speed. In $\texttt{evaluated}$ mode, each iteration requires approximately 1.5 seconds for the ray-tracing simulation and an additional 0.7 seconds for backpropagation using Dr.Jit, resulting in a total of ~2.2 seconds per iteration. Given convergence at approximately 38 iterations, the overall calibration time is roughly 88 seconds. Notably, 
although Sionna’s $\texttt{symbolic}$ mode can, in principle, execute a single iteration in approximately 0.1 seconds, this accelerated mode is not applicable during the calibration process, where the imposed evaluated-mode execution results in almost a 20-fold increase in per-iteration computation time. In stark contrast, the \textbf{\texttt{GAN MaxiViT-M2M}} model takes only 0.05 seconds for its inference, whereas  for \textbf{\texttt{GAN MaxiViT-C}} the execution time for one iteration is  0.13 ms (0.05 for inference plus 0.08 seconds for backpropagation, respectively) and converges in around 60 iterations. This corresponds to an effective speed-up of approximately 10-fold relative to the differentiable ray-tracing calibration process. This substantial difference in calibration time can also be observed in  Figure~\ref{fig:Time_Calibration_Result_Comparison},   which depicts the distribution of convergence times across models, together with an inset plot for \textbf{\texttt{GAN MaxViT-C}}, highlighting its faster convergence.


\begin{tcolorbox}[maintakeaway]
\textit{\textbf{Key insights:}} Differentiable ray-tracing exhibited substantial challenges in scaling to production-grade network setups,  both in terms of accuracy and speed.  On the other hand, pre-trained DL models with MNO data can attain out-of-area commensurate accuracy levels, whilst when undergoing an equivalent site-specific calibration process, they converge 10 times faster and are more accurate by roughly 3 dB. Consequently, we find substantial benefits in DL-based radio map generation, though only on the basis that they use a large volume of real world data.
\end{tcolorbox}

\section{Downstream Tasks and Implications}
\label{sec:downstream_tasks}

Beyond the differences in fidelity across the radio propagation models evaluated in Section~\ref{sec:models-comparison}, it is crucial to assess their implications for downstream tasks in wireless networks. To this end, we focus on two representative use cases---power saving optimization \cite{Gabrielle_Paper} and handover optimization \cite{kalntis_infocom25}---and examine how different radio propagation models affect their final performance.

\subsection{Power saving optimization}  

We replicate the methodology proposed in \cite{Gabrielle_Paper} to investigate how channel gain estimates from different propagation models can be utilized to minimize the antenna transmission power in 
the RAN, subject to the throughput constraints of individual users across space. To generate realistic throughput demands, we rely on the open-source models of \cite{zanella2023characterizing}, which are derived from session-level application traffic traces obtained from real-world measurements. This task considers two optimization parameters: the bandwidth allocated to each user, which should be sufficient to meet their throughput requirements, and the transmit powers of the antennas.

We focus our evaluation on a dense urban subarea within the capital city (Area 6), that comprises five antennas operating in the 1,800~MHz band. For these antennas, we generate user throughput demands following \cite{zanella2023characterizing}, as well as radio maps obtained from \textbf{\texttt{Sionna}}, \textbf{\texttt{GAN MaxViT}}, and crowdsourced measurements (\textbf{\texttt{Measured Data}}). Then, we run the optimization solution with channel gain estimates from the three different radio maps, and compare the final performance results taking crowdsourced measurements as a ground truth. In general, we would expect inaccurate channel gain estimates lead to wrong SINR calculations, which may potentially end up in suboptimal transmission power and, in extreme scenarios, preventing the optimizer from finding a valid solution that satisfies user throughput requirements. 
The mathematical formulation of the optimization problem can be found in Appendix~\ref{sec:Appendix_Tasks_Power_Opt}, and further details about the proposed solution can be found in \cite{Gabrielle_Paper}.

\begin{figure}
    \centering
    \includegraphics[width=1\columnwidth,trim={0 0 0 5pt},clip]{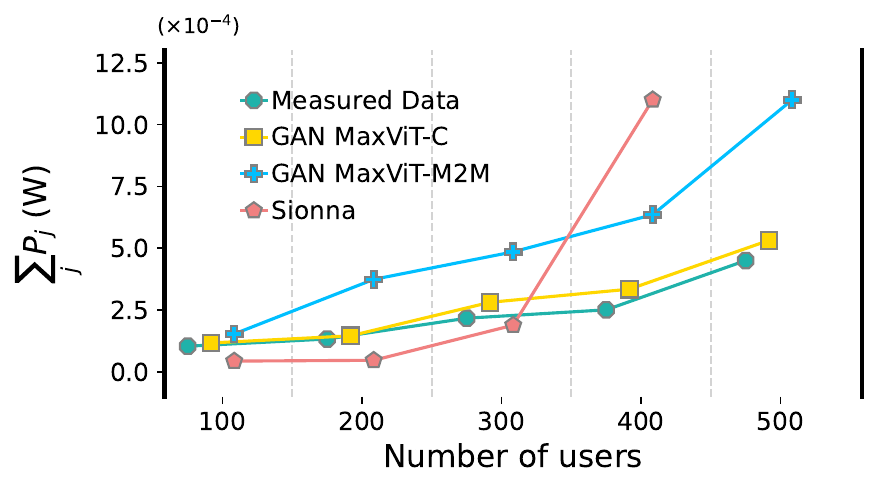}
    \caption{Power consumption across optimization instances under varying user load.}
    \label{fig:PowerOPT}
    \vspace*{-12pt}
\end{figure}

For our evaluation, we run the optimization algorithm over scenarios with an increasing number of users (100-500 users). Figure \ref{fig:PowerOPT} shows the total transmission power produced by each of the methods. The optimizer relying on Sionna yields a good alignment for up to 300 users, underpredicting, though, the power required by the system. As the number of users grows beyond 300, the optimizer starts to degrade compared to its counterparts relying of DL methods. Indeed, for 400 users, we observe an exponential increase of the consumed power, whilst for a higher number of users, the optimization algorithm could not attain an optimal solution. \textbf{\texttt{GAN MaxViT-M2M}}  follows the trend of the measured data but systematically overpredicts the required power, with the deviation becoming more evident as the number of users increases. Underestimating the required power may result in unmet user throughput demands in real-world deployments, whereas overestimation directly translates into unnecessary energy expenditures.
In contrast, \textbf{\texttt{GAN MaxViT-C}} demonstrates a closer alignment with the measured data across all evaluated scenarios, thereby providing a more reliable and cost-effective surrogate for power optimization.

\jsv{Can we add the improvement (in \%) between MaxVIT and Sionna for 400 users?}\jsv{How is it possible that Sionna performs better than the one based on the GT Measured Data? Are you measuring the results wrt the GT always?}





\subsection{Smooth Handover Handling}
Our second downstream task addresses handover optimization in scenarios with heterogeneous users exhibiting diverse mobility patterns. Specifically, we replicate the methodology presented in  \cite{kalntis_infocom25}, modelling user-antenna associations as online decisions considering users' mobility. The optimization problem jointly considers throughput maximization, determined by SINR estimates, and handover delay minimization, captured through a weighted switching penalty parameterized by user- and antenna-specific features. More details on the algorithm can be found in Appendix~\ref{sec:Appendix_Handovers}.

Similar to the previous task, we run the handover optimization algorithm~\cite{kalntis_infocom25} using signal strength estimates from \textbf{\texttt{Sionna}}, \textbf{\texttt{GAN MaxViT}}, and crowdsourced measurements (\textbf{\texttt{Measured Data}}), and use the latter as a reference to evaluate the final performance metrics after optimization. We consider a subarea within the capital city with five antennas operating on the 2,300 MHz band and distribute 1,000 users randomly at different locations. These 1,000 users are associated with four mobility profiles: static, pedestrian, cyclist, and vehicle, with probabilities [0.1, 0.1, 0.3, 0.5], respectively. Note that SINR changes more abruptly for users with higher-speed mobility patterns, thereby increasing handover frequency (i.e., higher cumulative switching penalty). More details on the user mobility profiles can be found in Appendix~\ref{sec:Appendix_Handovers}. Each optimization experiment runs for 5,000 timesteps, and between timesteps users move according to their mobility patterns. When radio maps lack channel estimates for a given location (\ie in \textbf{\texttt{Sionna}} and \textbf{\texttt{Measured Data}}), the location is mapped to the nearest point with available measurements, and the simulation continues.

\begin{figure}[!t]
    \centering
    \begin{subfigure}[b]{0.85\linewidth}
        \includegraphics[width=\linewidth]{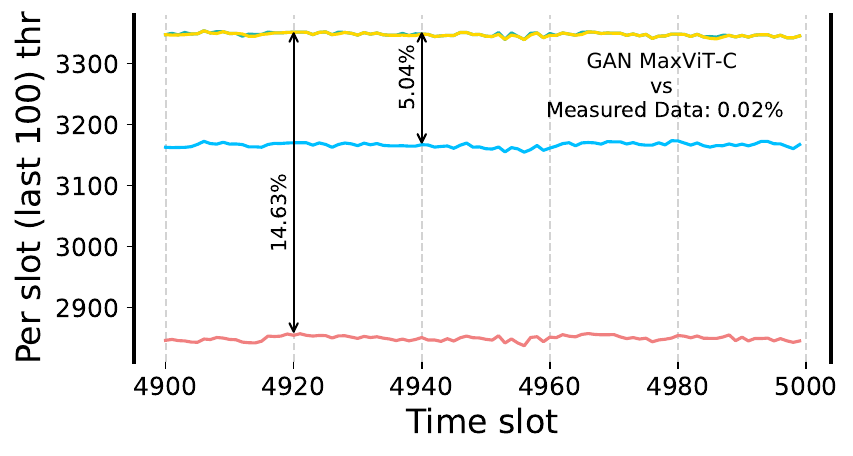}
        \caption{}
        \label{fig:Handovers_thr}
    \end{subfigure}
    \begin{subfigure}[b]{0.85\linewidth}
        \includegraphics[width=\linewidth]{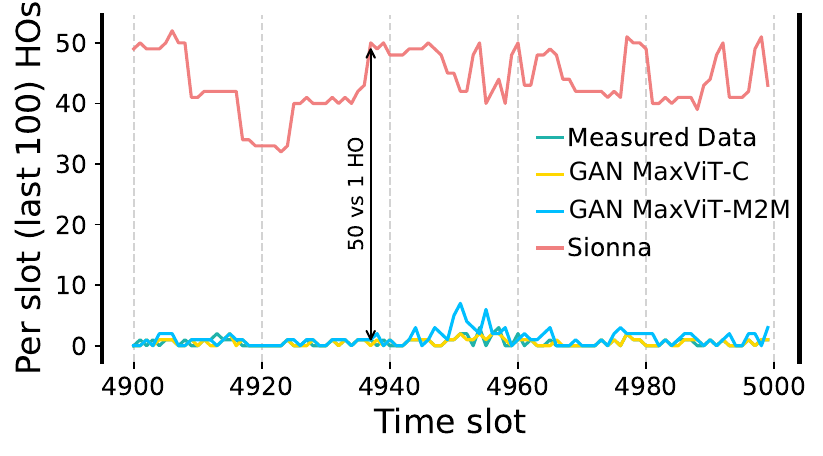}
        \caption{}
        \label{fig:Handovers_num}
    \end{subfigure}    
    \caption{Smooth handover handling: (a) Throughput and (b) number of handovers in the last 100 timeslots.}
    \label{fig:Handovers_thr_num}
\end{figure}


Figure \ref{fig:Handovers_thr_num} show that \textbf{\texttt{GAN MaxViT-C}} tracks the ground truth measurements (\textbf{\texttt{Measured Data}}) remarkably well, with the throughput deviating by only 0.02\% over the last 100 timeslots, compared to more than 14\% for \textbf{\texttt{Sionna}}. At the same time, the number of handovers differs substantially: ray-tracing estimates result in 40--50 handovers per timeslot (over 1,000 users in total), which is considerably higher than the number observed for the optimizers based on \textbf{\texttt{GAN MaxViT-M2M}} and \textbf{\texttt{GAN MaxViT-C}}. Overall, the results show that estimates from the calibrated DL-based model (\textbf{\texttt{GAN MaxViT-C}}) achieve performance comparable to the optimizer based directly on real-world measurements (\textbf{\texttt{Measured Data}}), while the DL model without site-specific calibration (\textbf{\texttt{MaxViT}}) still delivers highly competitive performance (-5.04\% throughput).

\section{Conclusion}

To deep learn, as things currently stand, that is the answer. 
Although differentiable ray-tracing introduced a leap forward in radio propagation modelling and demonstrated unprecedented simulation speed levels, it still faces significant challenges in scaling to real-world environments. Through a comprehensive large-scale experimental analysis using real-world data from a major MNO, our study highlights key limitations and demonstrates shortcomings relative to conventional DL models.
While ray-tracing can be of great benefit in scenarios without access to real-world measurements, DL models
%
demonstrated robustness, flexibility and the potential to generalize and generate high-fidelity radio maps in a variety of settings,
and are preferable when large volumes of data are available.
These findings are of paramount importance for the adoption of new tools and technologies for the emulation of RAN performance by MNOs, as well as for the scientific community and the course of future research.

\section*{Acknowledgement}

The authors acknowledge that differentiability is an indispensable component of all deep learning frameworks, and that the title may be technically imprecise. However, the title wording was chosen deliberately to preserve the literary structure of Shakespeare's original phrase.

\printbibliography

%

\appendix


\section{Evaluation metrics} 

This appendix provides the mathematical formulas for computing the error metrics used in Sections~\ref{Sec:R2R}–\ref{Sec:M2M}.

\label{sec:Appendix_Metrics}
\subsection{Full-Map Error Metrics}  \label{sec:Appendix_Metrics_R2R}

This set of metrics was used to evaluate the fidelity of \textbf{\texttt{R2R}} radio map reconstruction. Let \(\hat{y}_{ij}\) and \(y_{ij}\) denote the predicted and ground truth values at spatial position \((i, j)\), where \(i = 1, \dots, H\), \(j = 1, \dots, W\), with \(H = W = 512\). The RMSE, MAE, SSIM,  SMAPE, and PCC are computed as follows :

\begin{equation}
        \mathrm{RMSE} = 
        \sqrt{\frac{1}{NHW} 
        \sum_{n=1}^{N}\sum_{i=1}^{H} \sum_{j=1}^{W} 
        \left(\hat{y}_{n,ij} - y_{n,ij}\right)^2}.
    \end{equation}

    \begin{equation}
        \mathrm{MAE} = 
        \frac{1}{NHW} 
        \sum_{n=1}^{N}\sum_{i=1}^{H}\sum_{j=1}^{W} 
        \left|\hat{y}_{n,ij} - y_{n,ij}\right|.
    \end{equation}

 \begin{equation}
        \mathrm{SSIM}(x, y) = 
        \frac{(2\mu_x \mu_y + C_1)(2\sigma_{xy} + C_2)}
        {(\mu_x^2 + \mu_y^2 + C_1)(\sigma_x^2 + \sigma_y^2 + C_2)},
    \end{equation}

    \begin{equation}
        \mathrm{SMAPE} = 
        \frac{1}{NHW} 
        \sum_{n=1}^{N}\sum_{i=1}^{H}\sum_{j=1}^{W} 
        \frac{\left|\hat{y}_{n,ij} - y_{n,ij}\right|}
        {\left|\hat{y}_{n,ij}\right| + \left|y_{n,ij}\right| + \epsilon},
    \end{equation}
    
  \begin{equation}
        \mathrm{PCC} = 
        \frac{\sum_{n,i,j} 
        \left(\hat{y}_{n,ij} - \bar{\hat{y}}\right)
        \left(y_{n,ij} - \bar{y}\right)}
        {\sqrt{\sum_{n,i,j} 
        \left(\hat{y}_{n,ij} - \bar{\hat{y}}\right)^2}
        \sqrt{\sum_{n,i,j} 
        \left(y_{n,ij} - \bar{y}\right)^2}},
    \end{equation}

\noindent where \(\bar{\hat{y}}\) and \(\bar{y}\) are the global means of predictions and ground truth over all batches, \(\mu_x, \mu_y\) are patch means, \(\sigma_x^2, \sigma_y^2\) are patch variances, 
    \(\sigma_{xy}\) is the covariance, and \(C_1, C_2 > 0\) are stability constants, and \(\epsilon > 0\) avoids division by zero.

\subsection{Sparse Map Error Metrics}  \label{sec:Appendix_Metrics_R2M}


This set of metrics was used to evaluate the fidelity of \textbf{\texttt{R2M}} radio map reconstruction. For each sample \(n\),  we define the valid (non-zero) index set
\[
\mathcal{V}_n \;=\; \{(i,j)\,:\; y_{n,ij}\neq 0\},
\qquad
M \;=\; \sum_{n=1}^{N} |\mathcal{V}_n|
\]
as the total number of valid pixels across the batch. The metrics below are computed only over these valid indices.
 
\begin{equation}
\mathrm{RMSE}
\;=\;
\sqrt{\frac{1}{M}\sum_{n=1}^{N}\;\sum_{(i,j)\in \mathcal{V}_n}
\big(\hat{y}_{n,ij}-y_{n,ij}\big)^2 }.
\end{equation}
 
\begin{equation}
\mathrm{MAE}
\;=\;
\frac{1}{M}\sum_{n=1}^{N}\;\sum_{(i,j)\in \mathcal{V}_n}
\big|\hat{y}_{n,ij}-y_{n,ij}\big|.
\end{equation}
 
\begin{equation}
\mathrm{SMAPE}
\;=\;
\frac{1}{M}\sum_{n=1}^{N}\;\sum_{(i,j)\in \mathcal{V}_n}
\frac{\big|\hat{y}_{n,ij}-y_{n,ij}\big|}
{\big|\hat{y}_{n,ij}\big|+\big|y_{n,ij}\big|+\epsilon},
\end{equation}

\begin{equation}
\mathrm{PCC}
\;=\;
\frac{\displaystyle \sum_{n=1}^{N}\;\sum_{(i,j)\in \mathcal{V}_n}
\big(\hat{y}_{n,ij}-\bar{\hat{y}}\big)\big(y_{n,ij}-\bar{y}\big)}
{\sqrt{\displaystyle \sum_{n=1}^{N}\;\sum_{(i,j)\in \mathcal{V}_n}\big(\hat{y}_{n,ij}-\bar{\hat{y}}\big)^2}
\;\sqrt{\displaystyle \sum_{n=1}^{N}\;\sum_{(i,j)\in \mathcal{V}_n}\big(y_{n,ij}-\bar{y}\big)^2}}.
\end{equation}

\noindent where  $\bar{\hat{y}}$ and $\bar{y}$ are the means over the valid set: 
\[
\bar{\hat{y}} \;=\; \frac{1}{M}\sum_{n=1}^{N}\;\sum_{(i,j)\in \mathcal{V}_n}\hat{y}_{n,ij},
\qquad
\bar{y} \;=\; \frac{1}{M}\sum_{n=1}^{N}\;\sum_{(i,j)\in \mathcal{V}_n}y_{n,ij}.
\]

\subsection{Per Site MAE}  

This metric was used to assess the results of the site-specific radio map calibration. For each scene \(s\), we compute the mean absolute error (MAE) over all 2D spatial validation points \((i,j) \in \mathcal{V}_n\) at each optimization iteration \(k \in \mathcal{I}_n\), and select the minimum value achieved across iterations. Formally, this is expressed as

\begin{equation}
\text{MAE}_{\text{scene}, n} = \min_{k \in \mathcal{I}_n} 
\frac{1}{|\mathcal{V}_n|} \sum_{(i,j) \in \mathcal{V}_n} \big| \hat{y}_{ij}^{(k)} - y_{ij} \big|,
\end{equation}

\noindent where \(\hat{y}_{ij}^{(k)}\) denotes the predicted value at spatial location \((i,j)\) during iteration \(k\), and \(y_{ij}\) is the corresponding reference value. This formulation captures the best achievable validation MAE for each scene, accounting for all spatial locations for which we have measurements.

\section{Sionna Vectorized Calibration Formulation} \label{sec:Appendix_Sionna_vectorized}

To increase the representation capacity of each optimization variable, a vectorized parametrization formulation was proposed in \cite{SIonna_Learning}. Specifically, a \textit{read-out vector} projects a learnable \textit{embedding vector} onto a scalar value, thus allowing each optimization variable to be expressed as a flexible, continuous function of its underlying embedding. This approach enables richer parameter representations and improves the model’s ability to capture complex dependencies among variables. Each optimization parameter, $p$, is assigned a distinct \textit{embedding vector}, $w_p$, and two shared \textit{read-out vectors} are used for the different families of optimization parameters: one for the antenna parameters, $v_{ant}$, and one for the material parameters, $v_{mat}$.

\subsection{Vectorized Material Properties}  \label{sec:Appendix_Sionna_vectorized_Mat}

For the  vectorized material parameter representation, we adopt the same methodology employed in \cite{SIonna_Learning}, expressing the permittivity and conductivity of each distinct material as: 
 
\begin{equation}
\varepsilon_r = \exp\left( \frac{\mathbf{u}_{\mathrm{mat}}^\top \mathbf{w}_{\varepsilon}}{\sqrt{d}} \right) + 1,
\label{eq:material_permittivity}
\end{equation}

\begin{equation}
\sigma = \exp\left( \frac{\mathbf{u}_{\mathrm{mat}}^\top \mathbf{w}_{\sigma}}{\sqrt{d}} \right),
\label{eq:material_conductivity}
\end{equation}
\noindent where $\varepsilon_r$ denotes the relative permittivity and $\sigma$ denotes the conductivity of the material. 
The exponential parameterization ensures positivity, while the constant offset of $1$ in \eqref{eq:material_permittivity} 
enforces $\varepsilon_r \geq 1$, consistent with physical constraints.

\subsection{Vectorized Antenna Properties}  \label{sec:Appendix_Sionna_vectorized_Ant}

We consider a trainable version of the  \texttt{tr38901} antenna radiation pattern. Specifically, the antenna gain is given by

\begin{equation}
\label{eq:element_gain}
G(\theta, \phi) = 10^{\frac{A_{\mathrm{tot}}}{10}},
\end{equation}

\noindent where the total gain in dB is given by:

\begin{equation}
\label{eq:total_gain}
A_{\mathrm{tot}} = -\min\left( -\big(A_v(\theta) + A_h(\phi)\big), A_m \right) + G_{\mathrm{max}},
\end{equation}

\noindent whilst the vertical and horizontal attenuation components are defined as:

\begin{equation}
\label{eq:vertical_attenuation}
A_v(\theta) = -\min\left( 12 \left(\frac{\theta - \left(\frac{\pi}{2} - \theta_0\right)}{\mathrm{HPBW}_v}\right)^2, \mathrm{SLA}_v \right),
\end{equation}
\begin{equation}
\label{eq:horizontal_attenuation}
A_h(\phi) = -\min\left( 12 \left(\frac{\phi}{\mathrm{HPBW}_h}\right)^2, A_m \right),
\end{equation}

 \noindent where $\mathrm{SLA}_v$ and $A_m$ are the vertical side-lobe attenuation and the front-to-back gain cap, respectively, both set to $30~\mathrm{dB}$ attenuation. The remaining parameters of the attenuation components, \ie the steering angle, $\theta_0$, the vertical and horizontal half-width bandwidth $\mathrm{HPBW}_{v,h}$, and the maximum antenna element gain, $G_{max}$ are represented using a vectorized embedding parametrization: 
 
\begin{equation}
\label{eq:theta0_param}
\theta_0 = \theta_{\min} + (\theta_{\max} - \theta_{\min}) 
\cdot  g\left( \frac{\mathbf{u}_{\mathrm{ant}}^\top \mathbf{w}_{\theta}}{\sqrt{d}} \right)
\end{equation}

\begin{equation}
\label{eq:hpbwv_param}
\mathrm{HPBW}_v = HPBW_{v,h}^{\min} + (HPBW_{v,h}^{\max} - HPBW_{v,h}^{\min}) 
\cdot g\left( \frac{\mathbf{u}_{\mathrm{ant}}^\top \mathbf{w}_{v,h}}{\sqrt{d}} \right)
\end{equation}

\begin{equation}
\label{eq:gmax_param}
G_{\mathrm{max}} = g_{\min} + (g_{\max} - g_{\min}) 
\cdot g\left( \frac{\mathbf{u}_{\mathrm{ant}}^\top \mathbf{w}_g}{\sqrt{d}} \right)
\end{equation}

\noindent where $\theta_0 \in [-\pi, \pi]$, $\mathrm{HPBW}_v \in [10^\circ, 120^\circ]$, 
$\mathrm{HPBW}_h \in [10^\circ, 120^\circ]$, and $G_{\mathrm{max}} \in [0, 30]~\mathrm{dBi}$, 
and $g(\cdot)$ is the sigmoid activation function:
\begin{equation}
\label{eq:sigmoid}
g(x) = \frac{1}{1 + e^{-x}}.
\end{equation}

\section{Downstream Task Details}

\subsection{Power Optimization Problem Formulation} \label{sec:Appendix_Tasks_Power_Opt}

The considered optimization framework aims at minimizing the total transmission power in Orthogonal Frequency-Division Multiplexing (OFDM) systems by properly allocating the working bandwidth of each user $i$ connected to an antenna $j$, while meeting individual users' throughput requirements (bits/s). Formally, assuming a number of $n$ users and $N$ antennas,  the problem can be expressed as: 

\begin{subequations}
\label{min_in}
    \begin{align}
        & \min_{x_{ij}, P_j} \hspace{0.5cm} \sum_{j=1}^{N} P_j \\ 
        \textrm{s.t.:} & \nonumber\\
        & \,x_{ij}\in[0,1], \hspace{0.47cm} i\in[n], \quad j\in[N],\\
        & \sum_{i=1}^{n} x_{ij} \leq 1, \quad j\in[N],\label{eq:ncic_c5}\\
        & 
 x_{ij} B_j \log_2(1+\text{SINR}_{ij}(\mathcal{P})) \geq t_i, \quad i\in[n],\label{eq:ncic_c6}
    \end{align}
\end{subequations}
where
\begin{itemize}
    \item[i)] $P_j$ is the transmission power (W) of one resource block (RB) in base station $j$. Let $\mathcal{P}  \doteq [P_1,\ldots,P_N]$.
    
    \item[ii)] $t_{i}$ is  throughput requirement, measured in bits/s, of user $i$.

    \item[iii)] $B_j$: is the total bandwidth (Hz) of base station $j$.
    
    \item[iv)] $x_{ij}$ are the resources of BS $j$ assigned to user $i$. Consequently,  $x_{ij}{B_{j}}$ can be interpreted as the working bandwidth assigned by base station $j$ to user $i$.

    \item[vii)] $S(\mathcal{P})$: Signal-to-Interference-Noise-Ratio, defined as:
    \begin{equation}\label{SINR}
       \text{SINR}_{ij}(\mathcal{P}) = \frac{P_jg_{ij}}{\sigma^{2} + \sum_{k \neq j}P_k g_{ik}}, 
    \end{equation}
 where $\sigma^{2}$ denotes the noise power. 

    \item[viii)] $g_{ij}$ and $g_{ik}$: channel-gains obtained with the propagation models. 
\end{itemize}

The latter makes evident how the     channel gain computed via various models can implicitly affect the optimization results, as they are directly involved in one of the problems constraint related to throughput estimation and assurance.  

The above problem is highly non-convex; hence, it is typically solved via iterative techniques (such as gradient descent combined with successive linearization, difference of convex functions), which cannot guarantee a global optimal solution. The solution employed in  \cite{Gabrielle_Paper}, and replicated here, applies a piecewise concave approximation of the Shannon-Hartley Theorem followed by suitable change of variables ($P_j = e^{q_j}$ and $x_{ij} = e^{q_{ij}}$),  converting the (19a) - (19d)  into a Geometric Programming (GP) formulation, which is known to be convex. The resulting GP can be easily solved with standard literature solvers, such as \cite{mosek}, which we used in our experiments. Specifically, the GP formulation reads: 
\begin{equation}\label{min_GP}
\begin{aligned}
 \min_{u_{ij}, q_j} \quad & \sum_{j=1}^{N} e^{q_j}\\
\text{s.t.: } & e^{q_j}  \leq P_{j}, \quad j\in[N],\\
  & e^{u_{ij}} \leq 1, \quad  i\in[n], j\in[N]\\
  & \sum_{i=1}^{n} e^{u_{ij}} \leq 1, \quad  j\in[N],\\ 
  & \hat{f_\ell}(q_j,u_{ij}) \leq \frac{\log\left(\frac{B_ja_\ell}{t_i}\right)}{b_\ell}, \quad \ell\in[m], i\in[n], j\in[N],\\ 
\end{aligned}
\end{equation}
with 
\begin{equation}
    \hat{f_\ell}(q_j,u_{ij}) \doteq \log \left(\frac{\sigma^{2}}{g_{ij}}e^{-q_j-\frac{u_{ij}}{b_\ell}}+\sum_{k \neq j} \frac{g_{ik}}{g_{ij}}e^{q_k-q_j-\frac{u_{ij}}{b_\ell}}\right).
\end{equation}
$a_\ell$ and $b_\ell$ are constants used in the different $\ell$ functions designed to formulate the piecewise concave approximation. Both optimization problem formulations clearly show that inaccurate channel gain estimates can lead to low SINR values. Consequently, in such cases, the users' throughput constraints become more difficult to satisfy, potentially rendering the problem intractable.

\subsection{Smooth Handover Handling Problem Formulation}~\label{sec:Appendix_Handovers}

The handover algorithm of \cite{kalntis_infocom25} considers a heterogeneous cellular network consisting of a set $\mathcal J$ of $J$ antennas that serve a set $\mathcal I$ of $I$ UEs. It further assumes that a central controller, such as the Radio Access Network Intelligent Controller (RIC) in O-RAN \cite{oran_polese}, takes decisions for multiple UEs/antennas in a time-slotted manner for a set $\mathcal T$ of $T$ timeslots. Let  $x_{ij}(t)\in\{0,1\}$ be the (binary) association of user $i$ with antenna $j$ at the beginning of the timeslot $t$, where $x_{ij}(t)=1$ if user $i$ is associated to antenna $j$ at timeslot $t$, and $x_{ij}(t)=0$, otherwise. For presentation purposes, a vector $\bm x_t\!=\!\big(x_{ij}(t) \!\in\! \{0,1\}, i\in\mathcal I, j\in\mathcal J\big)$ is also defined. Then, the problem the network controller aims at solving can be formulated as follows:
\begin{align}
 \mathbb{P}: \ & \notag \max_{\{\bm{x}_t\}_{t}} \bigg\{\sum_{t=1}^T\sum_{i=1}^I\sum_{j=1}^J x_{ij}(t) \log \frac{c_{ij}(t)}{y_j(t)} - \gamma \sum_{t=1}^T\|\bm{x}_t-\bm{x}_{t-1}\|_A \bigg\}\\
	\textrm{s.t.} & \sum_{j\in\mathcal J}x_{ij}(t)=1, \quad \ \ \ \forall i\in\mathcal I, t \in \mathcal T \label{problem-assignment-constr}\\
	&y_j(t)=\sum_{i\in \mathcal I} x_{ij}(t), \ \  \forall j\in \mathcal J, t \in \mathcal T \label{problem-load-constr}\\ 
    & c_{ij}(t) \!=\! B_j\log_2 \! \big(1+ \text{SINR}_{ij}(t)\big), \forall j \!\in\! \mathcal J, i \!\in\!\mathcal I, t \!\in\! \mathcal T \label{problem-thr}\\
	&\bm{x}_t\in\{0,1\}^{I\cdot J}, \ \ \ \quad \ \ \forall t \in \mathcal T. \label{problem-binary-constr}
\end{align}

\noindent where \eqref{problem-assignment-constr} ensures that each user is served by one antenna, \eqref{problem-load-constr} defines the load of the antenna as the total number of associated users, and \eqref{problem-thr} measures the throughput of each antenna, similar to \eqref{eq:ncic_c6}, but assuming the SINR can change in each timeslot. The first term of $\mathbb P$ captures the total throughput, where we assume the antenna resources are allocated fairly across the users, while the second term determines the handover cost through the association change $x_{ij}{(t)} \neq x_{ij}{(t-1)}$, weighted differently for each user and antenna through the matrix $A$. Finally, $\gamma$ can be used as a scalarization and/or prioritization parameter. 

Solving $\mathbb P$ offline is impossible since at the beginning of the timeslots, the controller lacks knowledge of the future SINRs; it has no access or control over how users will move in the future. For this reason, the problem must be solved online. In addition, the problem is further compounded by the discreteness of the association decisions $\bm x_t$, and the handovers introduce a memory effect (of one timeslot), where past decisions affect the current.

To solve this, the authors propose a meta-learning algorithm based on the theory of Online Convex Optimization \cite{hazan-book}, which combines the decisions from a set of deployed experts (i.e., algorithms or learners), each with a different learning rate. In this way, even though SINRs (and thus, the best association decisions) can change significantly between timeslots, with careful selection of the different learning rates, at least one expert can perform well; and the goal of the meta-learner is to find and choose it. The interested reader is kindly referred to \cite{kalntis_infocom25} for additional information, which includes all the necessary details to reproduce the authors' work.

Regarding the mobility of these users, we adopt the Gauss-Markov model \cite{gauss_markov}, parametrized by velocities in meters per second (mps) and a randomness parameter $a \in [0,1]$, where $a = 0$ ($a = 1$) represents a totally random (linear) motion; determining, therefore, the memory and variability. Based on these parameters, the model predicts the future location, speed, and direction, creating realistic user movements. The parameters for the 4 considered profiles are the following: for static users, we assume velocities in [0, 0.2] mps and $a=0.99$, modeling near-constant positions; for pedestrian, velocities in [0.8, 1.6] mps and $a=0.8$, representing moderate randomness in their patterns; for cyclist, velocities in [4, 7] mps and $a=0.88$, reflecting relatively stable but faster mobility; and for vehicle, velocities in [10, 25] mps and $a=0.96$, representing high-speed but smoother trajectories.

To account for the switching costs expressed through handover delays, each profile is then mapped to a device type sampled from a empirical distribution from MNO's measured data over 31k real devices. We focus on user types associated with mobility, namely smartphones, tablets, wearables, IoT devices, and feature phones, which are captured with probabilities 93.5\%, 3\%, 1.1\%, 0.5\%, and 1.9\%, respectively. This distribution arises from our focus on users who execute horizontal (i.e., intra-LTE) handovers, where each device type is associated with distinct handover delays \cite{kalntis_infocom25}.

\end{document}